\documentclass[aps,reprint,twocolumn,floatfix,superscriptaddress]{revtex4-2}
\usepackage{xcolor}
\usepackage{graphicx}
\usepackage[] {hyperref}
\usepackage{physics}
\usepackage{comment}
\usepackage{CJKutf8}

\makeatother

\begin{document}
\begin{CJK}{UTF8}{gbsn}

\title{Entanglement phase transition due to reciprocity breaking without measurement or post-selection} 
\author{Gideon Lee} 
\affiliation{Pritzker School of Molecular Engineering, The University of Chicago,
Chicago, Illinois 60637, USA}
\author{Tony Jin} 
\affiliation{Pritzker School of Molecular Engineering, The University of Chicago,
Chicago, Illinois 60637, USA}
\author{Yu-Xin Wang (王语馨)}
\affiliation{Pritzker School of Molecular Engineering, The University of Chicago,
Chicago, Illinois 60637, USA}
\author{Alexander McDonald}
\affiliation{Institut Quantique \& D\'epartement de Physique, Universit\'e de
Sherbrooke, Sherbrooke, Qu\'ebec, J1K 2R1, Canada}
\author{Aashish Clerk}
\affiliation{Pritzker School of Molecular Engineering, The University of Chicago,
Chicago, Illinois 60637, USA}
\begin{abstract}
Despite its fully unitary dynamics, the bosonic Kitaev chain (BKC) displays key hallmarks of non-Hermitian physics including non-reciprocal transport and the non-Hermitian skin effect.  Here we demonstrate another remarkable phenomena: the existence of an entanglement phase transition (EPT) in a variant of the BKC that occurs as a function of a Hamiltonian parameter $g$, and which coincides with a transition from a reciprocal to a non-reciprocal phase.  As $g$ is reduced below a critical value,
the post-quench entanglement entropy of a subsystem of size $l$ goes from
a volume-law phase where it scales as $l$ to a \emph{super-volume
}law phase where it scales like $lN$ with $N$ the total system size.
This EPT occurs for a system undergoing purely unitary evolution and does not involve measurements, post-selection, disorder or dissipation.  We derive analytically the entanglement entropy out of and at the critical point for the $l=1$ and
$l/N\ll1$ case. 
\end{abstract}
\maketitle
\end{CJK}

\section{Introduction}

Recent years have seen intense efforts focused on understanding
entanglement dynamics in many-body quantum systems
with non-unitary evolution. It was found that for chaotic systems, measurements could trigger a novel phase transition from an entangled phase to a disentangled phase, a phenomenon dubbed a measurement-induced phase transition (MiPT) \citep{MiPTCaoTilloyDeLuca,Skinner_PRX_2019_MiPT,Li_PRB_2018_QZeno,Gullans_PRX_2021_quantum_coding_MiPT,Bao_PRB_2020_theory_of_MiPT,Choi_PRL_2020_Natural_QEC,Alberton_2021, Potter_MiPT_Review_2022, Fisher_MiPT_Review_2023}. While these models provide a fertile ground for the development of rich theoretical ideas at the intersection of statistical physics and quantum information \citep{BuchholdMiPTPRX,TurkeshiMiPTinfinitezeroclick,TurkeShiMiPThybrid,JinMartinMiPTclassical,NahumNLsM,LudwigNLsM,MirlinNLsM}, 
direct experimental observation of MiPTs can be extremely challenging, as this requires access to the full conditioned evolution
\cite{Minnich_2023_expt} (though note alternative strategies
based on quantities other than entanglement have been studied both theoretically \cite{Gullans_2020,Ippoliti_2021_xtdual,NaturePhyMiPTExp,Iadecola_2022,Buchhold_arXiv_2022_preselection,LiYd_2022,Dehghani_2022} and experimentally~\cite{NaturePhyMiPTExp,Google_2023_spacetime_duality_expt}). 

While MiPT is typically studied in systems where entanglement must be averaged over different random trajectories corresponding to distinct measurement outcomes, recent work has shown that entanglement phase transitions (EPT) can also occur without any stochasticity, in systems evolving under a non-Hermitian Hamiltonian \cite{LeGal_SciPost_2023_NH_MiPT_free_fermions, Kawabata_PRX_2023_EPT_from_NHSE, Gopalakrishnan_PRL_2021_NHEPT}.  
There is a direct connection to MiPT, as non-Hermitian dynamics is naturally interpreted as arising from measurement dynamics where one post-selects on a specific set of null measurement outcomes.  Of particular interest here are studies of EPT in non-Hermitian models exhibiting non-reciprocity (e.g.~directional systems where hopping to the right is much stronger than to the left).    
Kawabata et al.~\cite{Kawabata_PRX_2023_EPT_from_NHSE} studied an EPT in such a system (two coupled fermionic Hatano-Nelson~\cite{Hatano_1996_HNmodel,Hatano_1997_vortexpinning} chains) from volume law to area law entanglement scaling.  They found that the transition coincided with the transition between a reciprocal and a non-reciprocal phase.  The latter could be directly witnessed by the non-Hermitian skin effect (NHSE), a phenomenon occurring in the non-reciprocal phase where all modes localize under open boundary conditions
\citep{Yao_PRL_2018_NHSE_OG_2,Martinez_PRB_2018_NHSE_OG_3, Kunst_2018_PRL_biorth_correspondence,McDonald_PRX_2018_BKC,Okuma_2022_NH_review, Ashida_2020_NH_review, Bergholtz_2021_RMP_NH_review, Lin_2023_frontiers_topological_NH_review, Hatano_1996_HNmodel, Hatano_1997_vortexpinning}.
Despite this striking correspondence, one could still view the EPT transition here as being measurement driven, as the strength of non-reciprocity is directly tied to the strength of a post-selected measurement.  

Taking inspiration from these previous studies, in this work we ask whether an EPT can occur without any stochasticity {\it and} without any need for measurements (post-selected or not).  Similar to Ref.~\cite{Kawabata_PRX_2023_EPT_from_NHSE}, we consider a translationally-invariant model that exhibits a transition between reciprocal and non-reciprocal phases (with the latter exhibiting the NHSE).  In contrast to that work, our model has a {\it fully Hermitian} Hamiltonian and unitary evolution, and there is no need for any kind of measurement, post-selection or dissipation. 
Our setting is Hermitian, quadratic many-body bosonic Hamiltonians that do not conserve particle number.  Such models can exhibit non-reciprocity:  despite being fully unitary, the dynamics can nonetheless exhibit directionality at the level of the equations of motion for quadratures 
\citep{McDonald_PRX_2018_BKC,Wang_PRA_2019_NH_wo_dissipation,del_Pino_2022_NH_via_sq, Wanjura_2023_quadrature_nonreciprocity}.  Given that such models can exhibit non-reciprocity transitions, can they also exhibit entanglement transitions despite the lack of any connection to measurements?   
 

\begin{figure*}[hptb]
\centering \includegraphics[width=0.9\textwidth]{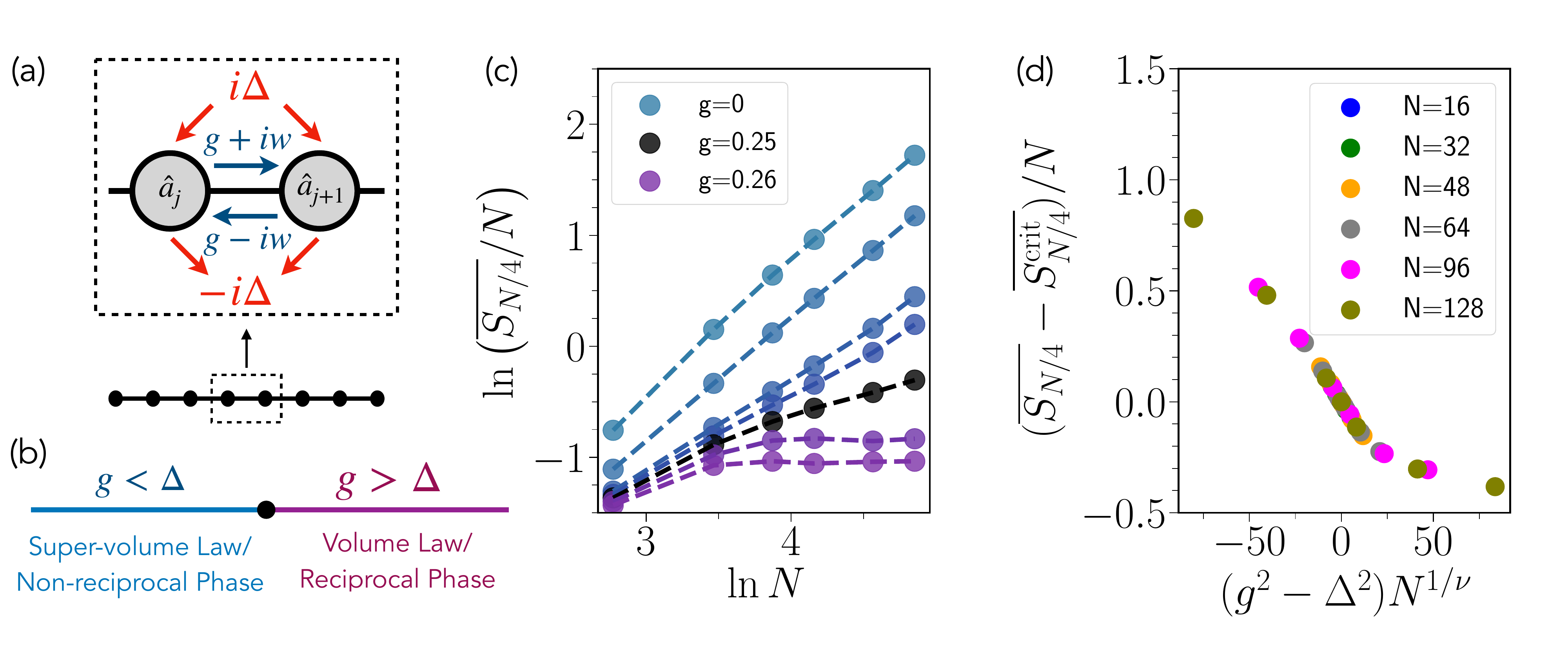} \caption{\textbf{a.} Schematic
of a 1D BKC lattice, with Hermitian hopping (amplitude $g + i w$), and 
pairing / two-mode squeezing (amplitude $i \Delta$) on each bond 
(see Eq.~(\ref{eq:H.BKC.gen})). \textbf{b.} Schematic phase diagram depicting super-volume law and volume law phases that arise as a function of $g / \Delta$ (with $g,\Delta < w$). 
For $g < \Delta$, the BKC exhibits the NHSE, and has nonreciprocal dynamics leading to super-volume law entanglement scaling (see main text), whereas for $g > \Delta$, the BKC is reciprocal, and exhibits the usual volume law of free systems. \textbf{c.}  Log-log plot of long-time-averaged EE for a subsystem
formed by the leftmost $N/4$ sites, divided by system
size. Circles are values from numerical simulation,
dashed lines are simply a guide to the eye. Since the
$y$ axis is divided by $N$, a slope of $0$ (greater than $0$) indicates volume law (super-volume law).  A clear transition is seen as $g$ is increased above $\Delta$.  
We fix $w = 1, \Delta = 0.25$, and the values of $g$ for the lines from blue to purple are $0, 0.2,0.24,0.245, 0.25, 0.255, 0.26$. 
\textbf{
d.} Scaling collapse of the long-time averaged EE for subsystem of
size $N/4$, with $\nu=0.5$ in Eq.~(\ref{eq:L4_scaling_collapse}).}
\label{fig:BKC_schematic} 
\end{figure*}

We find that the answer to this question is, surprisingly, yes. 
Our main result is to present the first instance of an EPT in a non-disordered bosonic system under purely unitary dynamics. We stress that this EPT requires \textit{no post-selection whatsoever}. Furthermore, as this model contains neither randomness nor measurements, one can unambiguously attribute this EPT to reciprocity breaking.   Our model of interest is a variant of the the bosonic Kitaev chain (BKC) model introduced in Ref.~\citep{McDonald_PRX_2018_BKC}.
By tuning a Hamiltonian parameter $g$, 
the BKC undergoes a phase transition from a phase where the dynamics of the $q$ and $p$ quadratures are non-reciprocal to a phase where they are reciprocal. At long times after a quench, the reciprocal phase is characterized by a volume law for the entanglement entropy (EE) of a subsystem of size $l$, i.e.~it scales linearly with $l$. On the other hand, the non-reciprocal phase has even stronger entanglement growth.  It is characterized by what we call a \emph{super-volume} law:  the EE of a subsystem of size $l$ scales as $lN$ where $N$ is the total system size.  Hence, if take a symmetric bipartition, the EE grows as $N^2$.   These behaviours are in stark contrast with \cite{Kawabata_PRX_2023_EPT_from_NHSE}, which also tied entanglement and reciprocity transitions, but found that non-reciprocity is detrimental to entanglement generation, leading to area law behaviour (see Sec.~\ref{sec:EPT}).  We also note that our results (for an unmeasured system) are distinct from the behaviour of explicitly monitored quadratic bosonic systems, which do not exhibit an EPT \cite{Zhou_PRB_2021_Gaussian_MiPT_nogo, Minoguchi_scipost_2022_bosonic_CFT}.  In addition to being of fundamental interest, our setup is also attractive for experiments.  Not requiring any measurements nor post-selection greatly simplifies implementation, while the the Hermitian bosonic pairing terms we require can be implemented in a variety of different platforms (they correspond to parametric drives or parametric down-conversion).      

The remainder of this paper is organized as follows.  In Sec.~\ref{sec:Model}, we recall the basic phenomenology of the BKC in both non-reciprocal and reciprocal phases. In Sec.~\ref{sec:EPT}, we present a numerical demonstration of the EPT. In Sec.~\ref{sec:SS_analysis}, we study analytically in depth the special minimal bipartition case $l=1$ and show that it captures already the essential features of the EPT. In Sec.~\ref{sec:GGE}, we extend the $l=1$ results to the case $l/N\ll1$ by relying on a local thermalization hypothesis towards a generalized Gibbs ensemble (GGE). Finally, in Sec.~\ref{sec:conclusion}, we conclude and discuss future directions.

\section{Model\label{sec:Model}}

The BKC describes bosonic modes on a $1D$ lattice that are coupled by hopping and pairing terms on each nearest-neighbor bond.  The Hamiltonian is 
\begin{equation}
\label{eq:H.BKC.gen}
\begin{aligned}\hat{H} & =\frac{1}{2}\sum_{j=1}^{N-1}\left((g+iw)\hat{a}_{j+1}^{\dag}\hat{a}_{j}+i\Delta\hat{a}_{j+1}^{\dag}\hat{a}_{j}^{\dag}+\text{H.c}.\right),\end{aligned}
\end{equation}
where $N$ is the total number of sites, $\hat{a}_{j}$ are bosonic operators, $[\hat{a}_{i},\hat{a}_{j}^{\dag}]=\delta_{ij}$, and $g, w$, $\Delta$ are real parameters of the model (see Fig.~\ref{fig:BKC_schematic}). We will call $\hat{\mathbf{r}}:=(\hat{q}_{1},\hat{p}_{1},...,\hat{q}_{N},\hat{p}_{N})^{T}$ the vector of quadrature operators, $\hat{q}_{j}:=(\hat{a}_{j}+\hat{a}_{j}^{\dag})/\sqrt{2},\hat{p}_{j}:=i(\hat{a}_{j}^{\dag}-\hat{a}_{j})/\sqrt{2}$.


Since $\hat{H}$ is quadratic, Gaussian states remain Gaussian under time evolution, and are fully specified by their 1-point function $\langle\hat{\mathbf{r}}\rangle$ and their $2N\times2N$ covariance matrix $\sigma_{ij}=\langle\{\hat{\mathbf{r}}_{i}-\langle\hat{\mathbf{r}}_{i}\rangle,\hat{\mathbf{r}}_{j}-\langle\hat{\mathbf{r}}_{j}\rangle\}\rangle$. In the remaining, we will fix the initial state to be the vacuum so that $\langle\hat{\mathbf{r}}\rangle=0$ at all times. The equations of motion (EOMs) for $\sigma$ close on themselves and are given by
\begin{equation}
\frac{d}{dt}\sigma=\Omega h\sigma+\sigma h\Omega^{T},
\end{equation}
where $h$ is the bosonic Bogoliubov-de Gennes (BdG) $2N\times2N$ matrix defined through $\hat{H}=\hat{\mathbf{r}}^{T}h\mathbf{\hat{r}}$ and $\Omega$ is the symplectic matrix $\Omega:=\bigoplus_{j=1}^{N}\begin{pmatrix}0 & 1\\
-1 & 0
\end{pmatrix}$. Note that the dynamics is completely linear in $\sigma$. 

The qualitative properties of the BKC are best understood by inspecting
the Heisenberg EOMs of $\hat{\mathbf{r}}$:
\begin{equation}\label{eq:quad_EOM}
\begin{aligned}\frac{d}{dt}\hat{q}_{j} & =\frac{w+\Delta}{2}\hat{q}_{j-1}-\frac{w-\Delta}{2}\hat{q}_{j+1}+\frac{g}{2}(\hat{p}_{j-1}+\hat{p}_{j+1}),\\
\frac{d}{dt}\hat{p}_{j} & =\frac{w-\Delta}{2}\hat{p}_{j-1}-\frac{w+\Delta}{2}\hat{p}_{j+1}-\frac{g}{2}(\hat{q}_{j-1}+\hat{q}_{j+1}).
\end{aligned}
\end{equation}
For $g=0$ (i.e.~purely imaginary hopping), these EOMs would describe independent,
non-reciprocal propagation of the $q$ and $p$ quadratures, with each having opposite directionality.  This mimicks the physics of two independent Hatano-Nelson chains~\cite{Hatano_1996_HNmodel,Hatano_1997_vortexpinning}. 

We focus throughout in this work on the case  $w>\Delta$ \footnote{For $w<\Delta$, both the OBC and the PBC case become unstable.}. In this regime, the model is always dynamically stable for open boundary conditions (OBC), while for
periodic boundary conditions (PBC) the system has a 
transition from stable to unstable as one goes from $\Delta<g$ to $g<\Delta$.
The stability of this bosonic system for $g=0$ can be conveniently understood in terms of amplification \cite{McDonald_PRX_2018_BKC}: starting from a wavepacket in the $q$ quadrature localized on one edge, the wavepacket will be amplified (damped) while propagating to the right
(left), and vice-versa for the $p$'s.
For PBC, this amplification
is unbounded, thus making the system unstable. In contrast, for OBC, the amplification
terminates at the edges, so that all average moments of the $q$'s are localized to the right and the $p$'s to the left. Adding a coupling $g$
mixes the quadratures together; as they have opposite directionality at $g=0$, this mixing diminishes non-reciprocity.  As $g$ increases, this mixing eventually prevents chiral amplification altogether,
leading to a sharp transition from a \textit{non-reciprocal} phase for $g<\Delta$ to a \textit{reciprocal} phase for $g>\Delta$. 

Another manifestation of this transition can be seen from the spectrum of the dynamical matrix $i \Omega h$ which exhibits the \emph{Non-Hermitian Skin Effect} (NHSE). This means that in the non-reciprocal phase, $g<\Delta$, for PBC the spectrum of $i \Omega h$ winds around $0$ in the complex plane, (i.e.~the system is unstable), whereas the spectrum under OBC collapses onto the real line (i.e., the system is stable). This is accompanied by the localization of all the OBC eigenmodes to the edges \cite{Yao_PRL_2018_NHSE_OG_2, Kawabata_PRX_2019_NHSE_Topology}. Conversely, in the reciprocal phase $g>\Delta$, the NHSE is absent -- the spectrum is always real, regardless of
boundary conditions. We note that the spectral properties of $i\Omega h$ are in complete analogy with the spectral properties of the non-Hermitian BdG Hamiltonian
presented in \cite{Kawabata_PRX_2023_EPT_from_NHSE} where an EPT was studied in coupled fermionic Hatano-Nelson chains. However, despite this similarity, we will show that the phenomenology in our Hermitian bosonic model is dramatically different.  Given our interest in phenomena induced by the NHSE, we will only consider OBC in the what follows.  

As shown in~\citep{McDonald_PRX_2018_BKC}, the BKC Hamiltonian $\hat{H}$ in Eq.~\eqref{eq:H.BKC.gen} can be mapped to a bosonic particle-conserving tight-binding chain via local unitary squeezing (i.e.~Bogoliubov) transformations. This effective local change of basis can be compactly written as 
$\hat{d}_{j} = e^{-i\phi j} \hat U _{j}
\hat{a}_{j} \hat U _{j} ^\dag $, with the unitaries given by
\begin{align}
\hat U _{j}
=  &
\begin{cases}
\hat S _{j} (r(j-j_0)) \hat S _{j} ( i r_{0}), 
& g<\Delta,
\\
\hat S _{j} ( i r_{0}), 
& g>\Delta,
\end{cases} 
\label{eq:diag_transform}
\end{align}
where $\hat S _{j} (\zeta) := 
e ^{ \frac{1}{2} ( \zeta \hat{a} ^{\dag 2}_{j} 
- \text{H.c.} ) }$ denotes the standard squeezing transformation on the $j$th site and $j_0$ is an arbitrarily fixed ``gauge'' parameter.

The parameters are given by $ r_0=\frac{1}{2}\tanh^{-1}
( {g}/{\Delta})$,
$r=\frac{1}{2}\log\frac{w+\sqrt{\Delta^{2}-g^{2}}}{w-\sqrt{\Delta^{2}-g^{2}}}$, $\phi=\frac{\pi}{2}$ for the non-reciprocal case $g<\Delta$, and $ r_0=\frac{1}{2}\tanh^{-1}({\Delta}/{g})$, $\phi 
= \arctan {(} w/
\sqrt{g^{2}-\Delta^{2}} )$ for the reciprocal phase $g>\Delta$.
After performing this transformation, the system is mapped in both cases to a simple tight-binding Hamiltonian 
\begin{equation}
\hat{H}= \sqrt{w^{2}+g^{2}-\Delta^{2}} \sum_{j=1}^{N-1}(\hat{d}_{j}^{\dag}\hat{d}_{j+1}+\text{H.c.}).
\end{equation}
The eigenmodes of $\hat{H}$ are standing waves, which we denote by 
$\hat{b}_n$ such that $\hat{H}=\sum_n \varepsilon_n\hat{b}^\dagger_n \hat{b}_n$ with $\varepsilon_n:=-2\sqrt{w^{2}+g^{2}-\Delta^2}\cos \left(\frac{\pi n}{N+1} \right) $.  

A few important points are in order.  First, note that as one increases $g$ across the non-reciprocity transition (i.e.~from below $\Delta$ to above $\Delta$), there is no signature of a transition in the eigenvalue spectrum:  the bandwidth of our system always increases monotonically with $g$.  Hence, the non-reciprocity to reciprocity transition we focus on {\it cannot} be simply diagnosed by looking at the spectrum of the OBC system.  Second, we stress that for both $g > \Delta$ and $g < \Delta$, our system has propagating quasiparticles.  In the non-reciprocal phase $g < \Delta$, it is useful to think of the position-dependent squeezing in Eq.~\eqref{eq:diag_transform} in terms of a localization length $\xi = a/r$, with $a$ the lattice spacing.  However, the emergence of this effective localization length does not impede quasiparticle propagation (i.e.~the group velocity remains finite).  
Finally, we note that to compute the EE of a subsystem of size $A$,
one can work either with the $\hat{a}_j$ or the $\hat{d}_j$ operators, as they are related to one another by purely local transformations.

\section{Entanglement phase transition\label{sec:EPT}}

We now turn to the study of the entanglement scaling across the different
phases. In \citep{Kawabata_PRX_2023_EPT_from_NHSE}, it was argued that non-reciprocity is detrimental for entanglement generation, as the quasi-particle pairs responsible for entanglement growth \citep{CalabreseCardyQuasiparticlepicture_2005} propagate in the same direction, thus preventing the generation of long-range correlations, and precluding any volume law scaling of EE in the non-reciprocal phase.  Here we show that the BKC, while presenting the main features of a non-Hermitian, non-reciprocal system (e.g.~the NHSE and non-reciprocal transport) deviates dramatically from this expectation. While the reciprocal phase indeed presents
a volume law as expected, we will show that the non-reciprocal phase
on the contrary fulfills a \emph{super-volume} law as defined in the introduction.

We consider the following setup. We fix the initial state to be the
physical vacuum (i.e. $\forall j, \hat{a}_j |\psi \rangle = 0$)
and allow it to evolve under the BKC Hamiltonian for a long time. Since the system is
dynamically stable, the time averaged EE of a subsystem will eventually converge
to some steady state value. For Gaussian states, the covariance matrix $\sigma$
fully determines the EE of any subsystem $\sigma$ \citep{RevModPhysGaussianQuantuminformation, HacklGaussianKahler,Serafini_2023_QCV_textbook}. The EE of a subsystem $A$ of size $l$ is
obtained from the relation 
\begin{equation}
S_{A}=\sum_{n=1}^{l}s(\nu_{n}),
\end{equation}
where 
\begin{equation}
    s(x):=\frac{x+1}{2}\ln\frac{x+1}{2}-\frac{x-1}{2}\ln\frac{x-1}{2}
    \label{eq:sDefinition}
\end{equation}
and $\nu_{n}$ are the positive eigenvalues of $i \Omega\sigma\vert_{A}$
where $\vert_{A}$ means that we truncate the support of the matrix
to $A$. Our quantity of interest will be the long-time averaged quantitiy $\overline{S}_A$, where we use the overline to denote time-averaged quantities. \begin{equation}
    \overline{f(t)} \equiv {\rm lim}_{T \rightarrow \infty} \frac{1}{T} \int_{0}^{T} f(t) dt.
\end{equation}

\begin{figure}[t!]
\centering \includegraphics[width=\columnwidth]{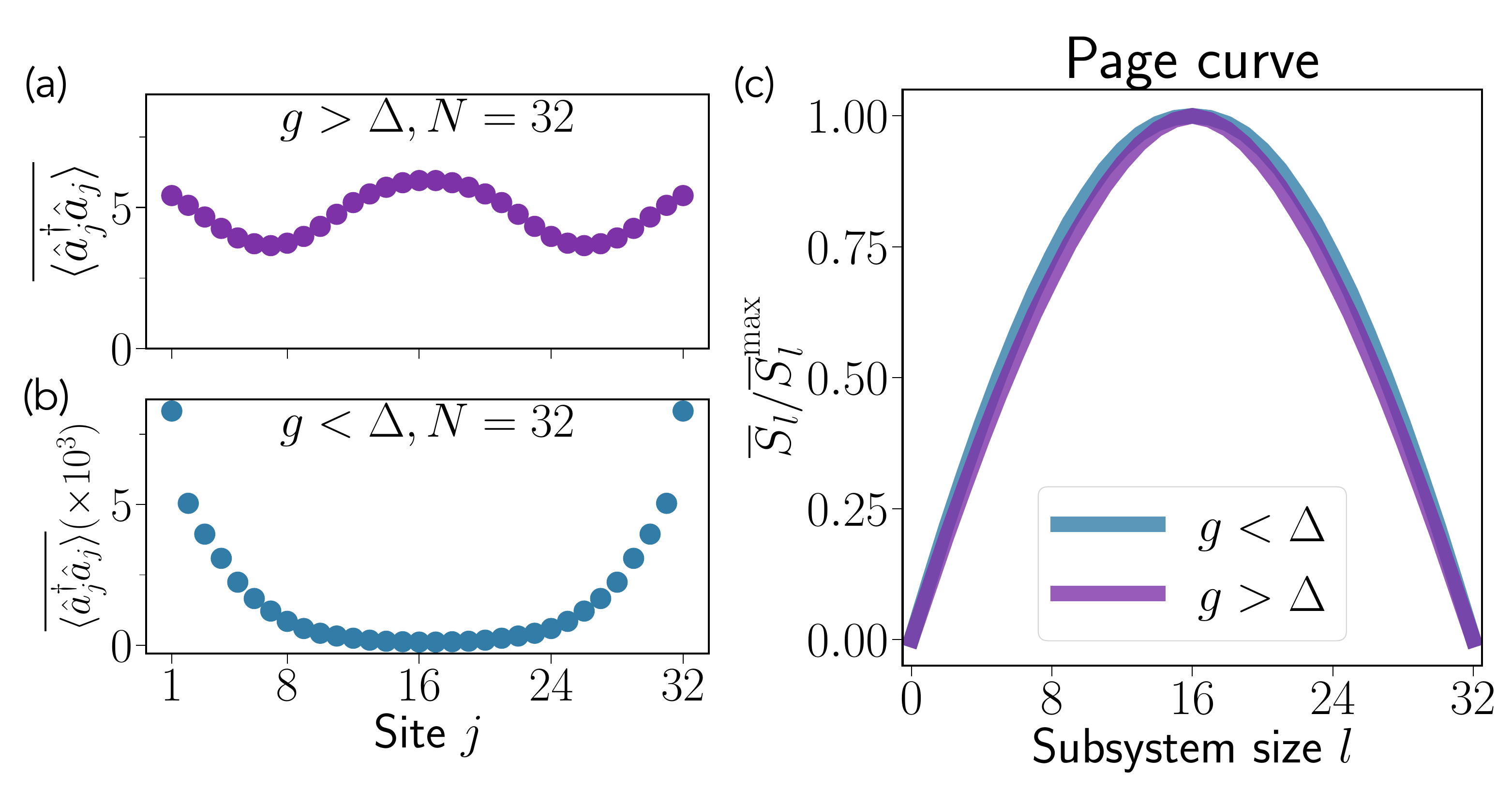} \caption{Comparison of the average density and Page curves for the reciprocal (with
$g=0.3$) and non-reciprocal (with $g=0.24$) phases, for $w=1,\Delta=0.25,N=32$.
\textbf{a.} Time-averaged position-dependent particle number $\overline{\langle\hat{a}_{j}^{\dag}\hat{a}_{j}\rangle}$,
in the reciprocal phase. \textbf{b.} Same as (a) but in the non-reciprocal phase. Note
the difference in shape and increase in scale in going from panel \textbf{a} 
to \textbf{b}.  (c) EE for a cut of size $l$, with
the subsystem being the left $l$ sites of the chain in both the non-reciprocal and reciprocal phases. The curves are normalized by their maximum value.
Despite the dramatic differences between (a) and (b), the Page curves are hardly distinguishable.}
\label{fig:occ_and_page} 
\end{figure}

Before examining the EE, we look at the time-averaged position-dependent particle density across
the chain. We observe that, as expected, the non-reciprocal phase
has particles exponentially localized to the edges (Fig.~\ref{fig:occ_and_page}a),
while the reciprocal phase does not. Thus, one may expect that, in
the reciprocal phase, taking a cut of the system from the left and
increasing its size will not lead to a significant increase of the
particle number and consequently, no significant increase in the
EE. However, this is not the case. To illustrate this fact, we plot
on Fig.~\ref{fig:occ_and_page}b the so-called Page curve \citep{PageCurve},
i.e the EE of a subsystem of size $l$ as a function of $l$ while
keeping the total system size $N$ fixed. One finds that the reciprocal
and non-reciprocal phases yield almost identically-shaped curves. We conclude, perhaps surprisingly, that the BKC does not show any area law phase induced by non-reciprocity.

\begin{figure}[tp]
\includegraphics[width=\columnwidth]{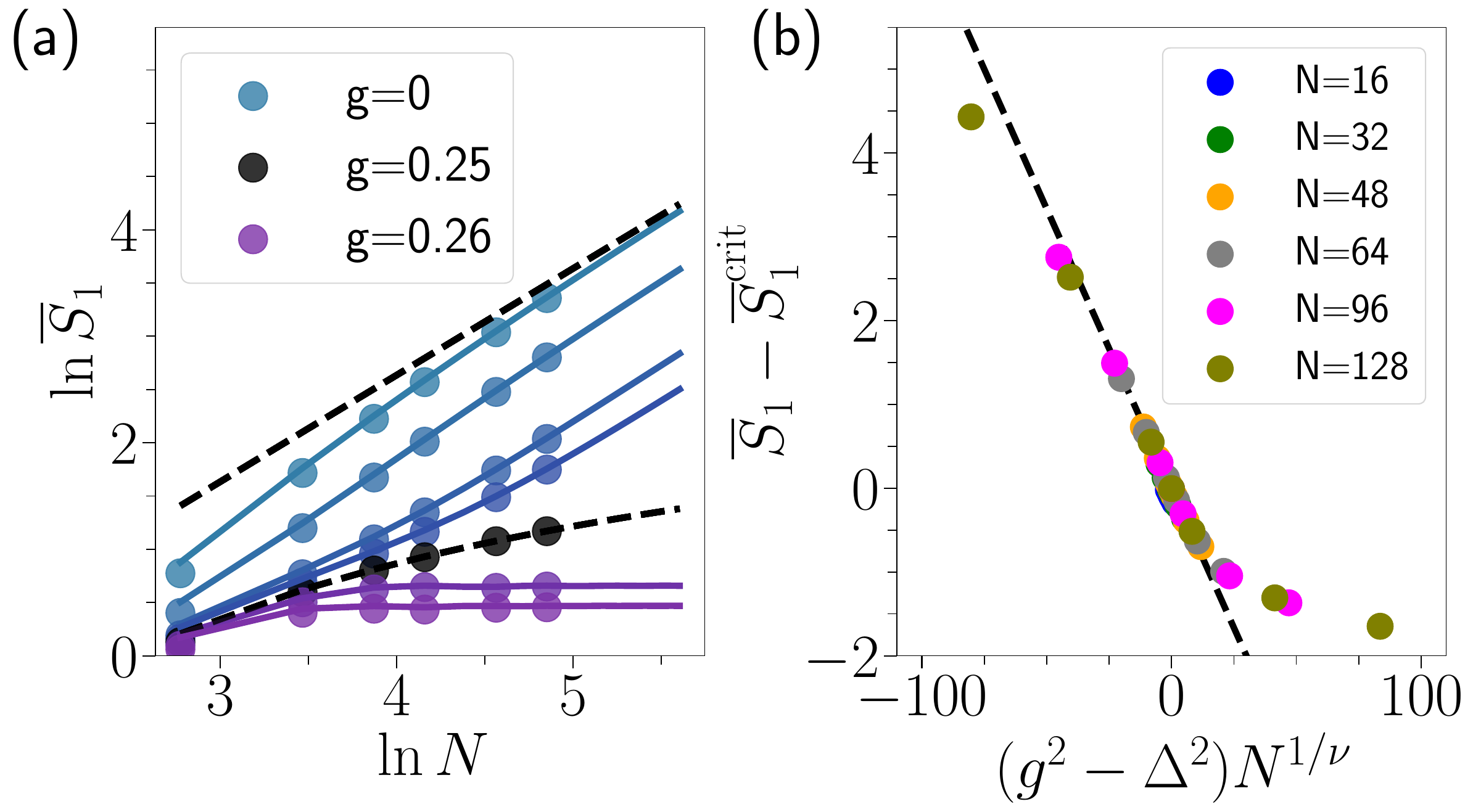}
\caption{\textbf{a.}~Log-log plot of long-time-averaged EE for the EE of the
left-most site $1$. The fixed parameters are $w = 1, \Delta = 0.25$. The values of $g$ for the lines from blue to purple are $0, 0.2,0.24,0.245, 0.25, 0.255, 0.26$. Scatter plots indicate values obtained from numerical
simulation, whereas the solid lines are the analytic prediction from Eqs.~(\ref{eq:eigenvaluenonrecp}) and (\ref{eq:eigenvaluerecp}). We observe
good agreement between the two for all but the smallest chain size.
The upper dashed line indicates the $Nr$ scaling in the non-reciprocal
phase, and the lower dashed line indicates the single-site EE at the
critical point $g=\Delta$ which separates the two phases. The two
distinct scalings, with $N$ in the non-reciprocal phase, and saturating
in the reciprocal phase, are readily apparent. \textbf{b.}~Scaling
collapse for single site EE with $\nu=0.5$ in Eq.~(\ref{eq:SSEE_scaling_collapse}). The black dashed line is the expansion near the critical point obtained from the analytical form of the entropy  (see Eqs.~(\ref{eqref:EEscalingcollapseSinglesite1}, \ref{eqref:EEscalingcollapseSinglesite2}), }
\label{fig:EE_numerics} 
\end{figure}

To more fully understand entanglement properties, we can also consider the scaling of subsystem EE in a slightly different manner.  Instead of fixing total system size $N$ and varying subsystem size, we can fix the subsystem
size $l$ to be a fraction of total system size, $l= z \times N$, and then vary the total
system size while keeping $z$ fixed.  For concreteness, we take $z=1/4$ in what follows.  The results for this scenario are presented in Fig.~\ref{fig:EE_numerics}\textbf{a}
for $w=1,\Delta=0.25$, and a range of $g$ close to $\Delta$. Usually, one does not have to worry about the total system size as long as it is large enough; however, here, we find drastically different phenomena. When scaling the total system size, we immediately observe the emergence of two
phases -- a ``super-volume'' law phase where the EE scales as $N^{2}$ corresponding to the non-reciprocal phase
$g<\Delta$ and a volume law phase where the EE scales as $N$ corresponding to the reciprocal phase $g>\Delta$. The two
are separated by a logarithmic scaling $N\log N$ when $g=\Delta$.

From these considerations and following the fitting procedure in e.g.~\cite{Skinner_PRX_2019_MiPT}, we attempt the following finite size scaling for the EE,
\begin{equation}
\overline{S_{N/4}(g,\Delta,N)}-\overline{S_{N/4}(\Delta,\Delta,N)}=Nf((g^{2}-\Delta^{2})N^{1/\nu})\label{eq:L4_scaling_collapse}
\end{equation}
and find that, fixing the critical exponent $\nu=0.5$, it yields a good quality collapse of the EE for different system sizes onto the same curve
, see Fig.~\ref{fig:EE_numerics}\textbf{b}. 

We thus have established a key result of our work:  despite the lack of measurements, postselection or disorder, our BKC model exhibits a clear entanglement phase transition as a function of $g$, one that coincides with the transition from a reciprocal to non-reciprocal phase.  

\section{Analytic proof of entanglement phase transition for a minimal bipartition\label{sec:SS_analysis}}

Computing the post-quench EE analytically, even for free systems, is in general a formidable
task \citep{XYspinChainEEcomputationFagotti,PNASQuasiparticleintegrablesytems,ParezEEExact}. In this section, we provide analytical insight by studying
the EE for the simpler \textit{minimal bipartition} case, i.e.~the case where the subsystem $A$ is composed of a single site.
We will see that a transition from a phase where the EE is $O(N)$ to a phase
where the EE is $O(1)$  already occurs for this case when increasing $g$ above $\Delta$. 

For a single site $j$, the instantaneous symplectic eigenvalue $\nu_{t}$ at time $t$
is given by 
\begin{equation}
\nu_{t}^2=\left(2\langle\hat{d}_{j}^{\dag}\hat{d}_{j}\rangle_t+1\right)^2-4|\langle\hat{d}_{j}\hat{d}_{j}\rangle_{t}|^2,\label{eqn:1site_eig} 
\end{equation}
where we recall that $\hat{d}$ refers to the tight-binding basis defined in Eq.~\eqref{eq:diag_transform}. The associated EE is simply $S_{1}=s(\nu_{t})$.
For now, we have not specified a particular site. Given the strong spatial non-uniformity in the density in the non-reciprocal phase (see Fig.~\ref{fig:occ_and_page}b), one would naturally expect that $S_1$ would depend strongly on the choice of site, with large values at the boundary, and small values in the middle of the chain.  Surprisingly, this is not the case. Eq.~\eqref{eq:diag_transform} already highlights how this can be.  It shows succinctly that the entanglement is not simply determined by average photon number alone: one also needs to understand how many of these photons are associated with purely local squeezing correlations, and separate out this contribution.  As we will see, despite the average density being highly inhomogeneous, $S_1$ is largely independent of the position $j$ of the chosen site.
   

\paragraph*{Non-reciprocal phase\label{ssec:NR_phase_SSEE}}

In the non-reciprocal phase, we expect $\nu_{t}$ to be exponentially
large with $N$, so that $S_{1}$ grows as $N$. When $\nu_{t}$ is large, the
EE takes a simple form: to leading order in $\nu_{t}$, $
S_{1}\approx\frac{1}{2}\ln\nu_{t}^{2}$. We will compute the time-average of the entropy, $\overline{S_1}$.
In general $\overline{\ln\nu_{t}^{2}}\neq\ln\overline{\nu_{t}^{2}}$ but it would be desirable to use the latter expression as it is much easier to compute. To quantify the error resulting from this approximation, we Taylor expand $\ln\nu_{t}^{2}$ with respect to $\nu_{t}^{2}$ around $\overline{\nu_{t}^{2}}$ and take the average to get :
\begin{equation}\label{eq:ln_conc}
\overline{\ln\nu_{t}^{2}}=\ln\overline{\nu_{t}^{2}}+\frac{\overline{(\nu_{t}^{2}-\overline{\nu_{t}^{2}})^{2}}}{2(\overline{\nu_{t}^{2}})^{2}}+...
\end{equation}
We show in App.~\ref{app:4pt} that the second term is of
order $O(1)$ because of cancellation of exponential-in-$N$ contributions to the numerator and denominator. Hence,
\begin{equation}
\overline{S_{1}}\simeq\frac{1}{2}\ln\overline{\nu_{t}^{2}}.
\end{equation}
Finally, to compute $\overline{\nu_t^2}$, we will make use of the fact that 

\begin{equation}\label{eq:4pt_surprise}
    \overline{\langle\hat{d}_{j}^{\dagger}\hat{d}_{j}\rangle^{2}}-\overline{|\langle\hat{d}_{j}\hat{d}_{j}\rangle|^{2}}=\left(\overline{\langle\hat{d}_{j}^{\dagger}\hat{d}_{j}\rangle}^{2}-\left |\overline{\langle\hat{d}_{j}\hat{d}_{j}\rangle}\right |^{2}\right) \left(1+O(1/N)\right),
\end{equation} 
 a property that we prove in App.~\ref{app:4pt}. 
 Remarkably, the neglect of fluctuations, implicit in this approximation, only holds for this difference, and not for each term individually.  More explicitly, one finds 
 $    \overline{\langle\hat{d}_{j}^{\dagger}\hat{d}_{j}\rangle^{2}} \neq \overline{\langle\hat{d}_{j}^{\dagger}\hat{d}_{j}\rangle}^{2},
$ and $\overline{|\langle\hat{d}_{j}\hat{d}_{j}\rangle|^{2}} \neq \left |\overline{\langle\hat{d}_{j}\hat{d}_{j}\rangle}\right |^{2}$. Our interpretation of this result is that while individual sites are subject to (exponentially) large temporal fluctuations in density, these fluctuations are almost entirely due to fluctuations in the amount of local pairing correlations.  The contributions of these fluctuations (density, local pairing) thus cancel each other to leading order when calculating the symplectic eigenvalue and, consequently, the EE.
Hence, to compute $\overline{\nu_t^2}$, one only needs the average covariance $\overline{\sigma}$.
Taken together, all these steps considerably simplify the task of computing the EE and allows us to have quantitative results.

To compute $\overline{\sigma}$, we will consider the continuum limit, which we define as follows : let $a$ be the lattice spacing, we consider the limit $N\to\infty$, $a\to0$, while keeping fixed the dimensionful quantities $\xi:=\frac{a}{r}$ (localization length), $L:=a(N+1)$ (system size), $x:=aj$ and $p:=\frac{\pi n}{a(N+1)}$. We are free to fix the parameter $x_{0}=aj_0$. For the particular choice of $x_{0}=\frac{L}{2}$, the correlations have a compact expression (see App.~\ref{App : Average correlations}): 
\begin{align}\overline{\langle\hat{d}_{x}^{\dagger}\hat{d}_{x}\rangle}&=\cosh(2 r_0)\frac{\sinh(L/\xi)}{2L/\xi}-\frac{1}{2}, \label{eq:localnormalcorrelationsNonrecp}\\
\overline{\langle\hat{d}_{x}\hat{d}_{x}\rangle}&=(-1)^{\frac{x}{a}}\frac{i}{2}\sinh(2 r_0).\label{eq:localanomalouscorrelationsNonrecp}
\end{align}
We see that the time-averaged local density and pairing correlations in the squeezed frame are \emph{spatially uniform}. At first glance, this could seem surprising, as in this frame, our initial condition (vacuum in the lab frame) is extremely non-uniform in space, due to the position dependent squeezing transformation in Eq.~\eqref{eq:diag_transform}.  
However, the resulting uniformity of the time-averaged state can be understood by the dynamics being equivalent to a simple tight-binding chain.  Indeed, for such a model, any spatial product state will lead to an average homogeneous profile in the continuum limit. 

This in turn means that the EE in the minimal bipartition protocol will yield the same result, independent of our choice of which site to single out.  Heuristically, this explains the discrepancy between Fig.~\ref{fig:occ_and_page}\textbf{b} and \textbf{c}: while the average density in the lab frame is exponentially localized towards the edges, this excess density can largely be attributed to local squeezing, which does not affect entanglement properties, in line with our interpretation of Eq.~(\ref{eq:4pt_surprise}) (see App.~\ref{app:sqplth_ansatz} for further discussion about the separation between local squeezing and thermal occupation). 
Inserting (\ref{eq:localnormalcorrelationsNonrecp},\ref{eq:localanomalouscorrelationsNonrecp}) in \eqref{eqn:1site_eig} leads to 
\begin{equation}
\overline{\nu_{t}^{2}}\approx1+\cosh^{2}\left(2 r_0\right)\left(\frac{\sinh^{2}(L/\xi)}{\left(L/\xi\right)^{2}}-1\right). \label{eq:eigenvaluenonrecp}
\end{equation}
Away from the critical point, the large
$L$ limit gives us that 
\begin{equation}
\overline{S}_{1}\approx L/\xi.
\end{equation}
To obtain the scaling near the critical point, we consider the limit $L/\xi\ll1$ for which
the localization length $\xi$ is far greater that the system size. This (see App.~\ref{sec:Equivalence-between-single}) leads to: 
\begin{equation}
\overline{S_{1}}\approx\ln N+\frac{1}{15}\frac{\Delta^{2}-g^{2}}{w^{2}}N^{2}. \label{eqref:EEscalingcollapseSinglesite1}
\end{equation}
\paragraph*{Reciprocal phase\label{ssec:RP_phase_SSEE}}
In the reciprocal phase, the local correlations are given in the continuum limit by  
\begin{align}
 & \langle\hat{d}_{x}^{\dagger}\hat{d}_{x}\rangle=\frac{1}{2}\left(\cosh\left(2 r_0\right)-1\right),\label{eq:localdensityrecp}\\
 & \overline{\langle\hat{d}_{x}\hat{d}_{x}\rangle}=(-1)^{x/a}\frac{\sinh\left(2 r_0\right)}{2}\frac{e^{i\varphi L}-1}{\varphi L}.\label{eq:localpairannihilationrecp}
\end{align} where we defined $\varphi:=\frac{1}{a}(\pi-2\phi)$. 
We find that the average density in the squeezed frame is both time-independent and spatially uniform.  This is no surprise:  in the reciprocal phase, the transformation to go to the squeezed frame (c.f. Eq.~\eqref{eq:diag_transform}) is uniform, hence our initial pre-quench state is also uniform.  Such a density profile will not evolve under a tight-binding Hamiltonian.  In contrast to this, the time-averaged local squeezing correlators above retain a position dependence in their phase.   

Away from the transition, i.e.~for $\varphi$ finite, the local pairing correlations tend to $0$ in the large $L$ limit. Thus the EE is simply 
\begin{equation}
S_{1}\approx s\left(\frac{g}{\sqrt{g^2-\Delta^2}}\right).
\end{equation}
where $s(x)$ is defined in Eq.~(\ref{eq:sDefinition}).
Close to the transition, the eigenvalue scales like $N$ and we can apply the same set of approximations used in the reciprocal case. The eigenvalue corresponding to (\ref{eq:localdensityrecp},\ref{eq:localpairannihilationrecp}) is 
\begin{equation}
\overline{\nu_t^2}\approx 1+\sinh^{2}(2 r_0)\left(\frac{2\left(1-\cos(\varphi L)\right)}{(\varphi L)^{2}}-1\right).\label{eq:eigenvaluerecp}
\end{equation}
Close to the critical point, the limit $\varphi L \ll 1$ leads to 
\begin{equation}
\overline{S_{1}}\approx\ln N+\frac{1}{15}\frac{\Delta^{2}-g^{2}}{w^{2}}N^{2}
\label{eqref:EEscalingcollapseSinglesite2}
\end{equation}
which is consistent with the limit Eq.~(\ref{eqref:EEscalingcollapseSinglesite1}) from the non-reciprocal phase. 
\paragraph*{Numerical simulation and scaling collapse}

Numerical simulations of the EE for a single-site
are plotted on Fig.~\ref{fig:EE_numerics}\textbf{a} and \textbf{b} alongside analytical estimates. We observe that for
large system size, the EE of the non-reciprocal phase always goes to the $L/\xi$ scaling, whereas the EE for the reciprocal phase saturates.
The expressions (\ref{eqref:EEscalingcollapseSinglesite1},\ref{eqref:EEscalingcollapseSinglesite2}) suggest the following scaling collapse
\begin{equation}
\overline{S_{1}(g,\Delta,N)}-\overline{S_{1}(\Delta,\Delta,N)}=f((g^{2}-\Delta^{2})N^{1/\nu}),\label{eq:SSEE_scaling_collapse}
\end{equation}
with $\nu=0.5$, thus proving Eq.~\eqref{eq:L4_scaling_collapse} for the minimal bipartition. Note that, not only the power laws are in agreement with the numerics but also the non-universal $1/15$ prefactor of the second term of (\ref{eqref:EEscalingcollapseSinglesite1},\ref{eqref:EEscalingcollapseSinglesite2}), see Fig.~\ref{fig:EE_numerics}.

\begin{figure}[tp]
\centering
\includegraphics[width=0.7\columnwidth]{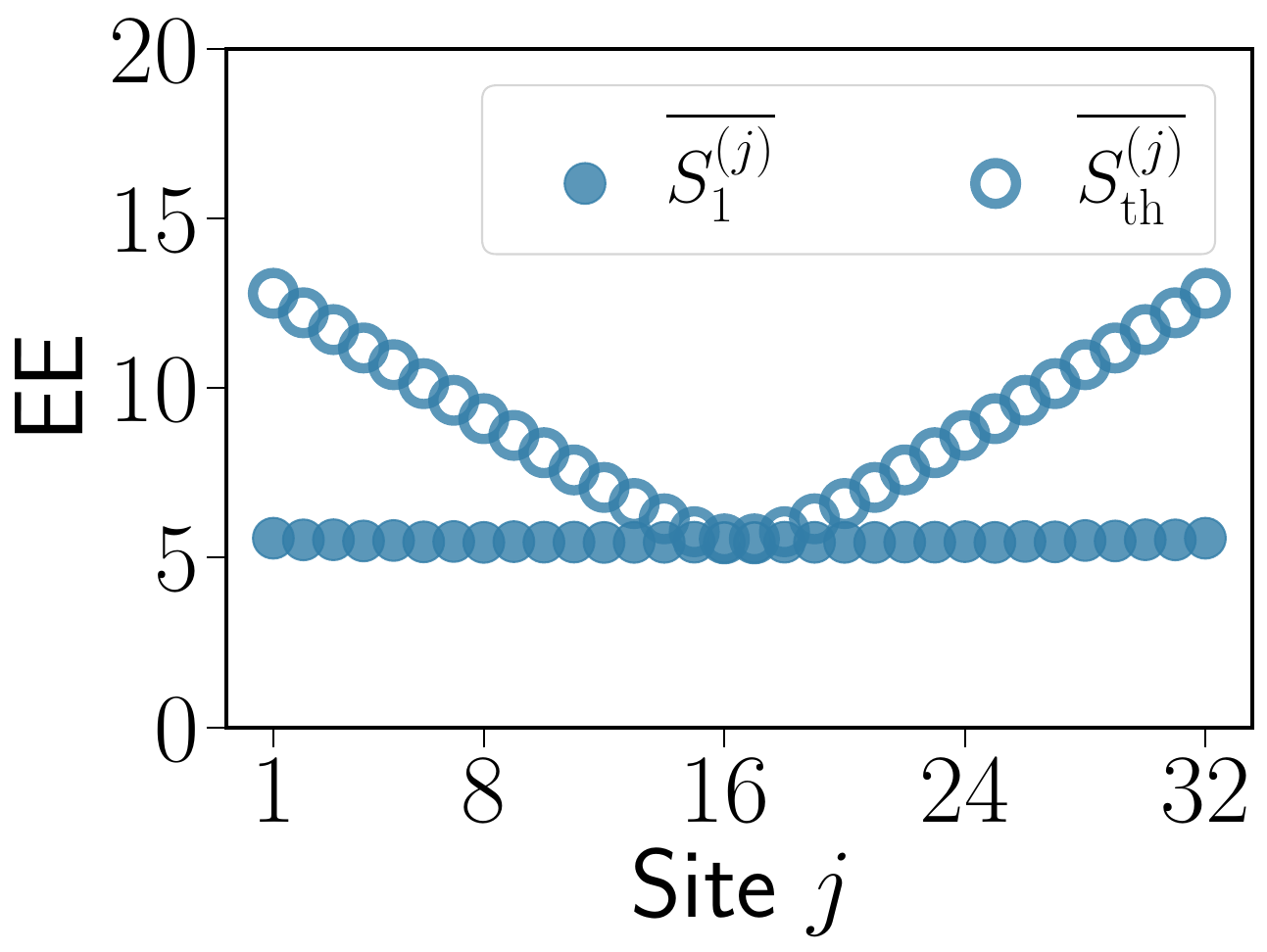}
\caption{
Filled circles: one-site entanglement entropy $\overline{S_1^{(j)}}$ as a function of position $j$ in the non-reciprocal phase, for parameters $N = 32, w = 1, g = 0, \Delta = 0.25$.  Despite being in the skin-effect phase, the entropy is almost completely uniform.  Open circles:  $\overline{S_{\rm th}^{(j)}}$, the prediction for the single site entropy based using the average photon number alone, c.f.~Eq.~\eqref{eq:S1thermal}.  This prediction deviates sharply from the true result.  This highlights an important caveat:  simply using average particle number as a proxy for entanglement can be extremely misleading.    }
\label{fig:thermal_mistake} 
\end{figure}

Before leaving this section, we wish to highlight a crucial fact: entanglement in our system {\it cannot} be simply predicted from the behaviour of average photon number.  A naive argument would be that the average photon number on each site of the post-quench state determines its effective Hilbert space dimension $D_j$, which would then (assuming thermalization) set its entropy.  This line of reasoning would suggest that the entropy of a given site should correspond to the entropy of a single bosonic mode in a thermal state, i.e. 
\begin{equation}
    \bar{S}^{(j)}_{1,{\rm th}} = 
    s \Bigg( \overline{  
    \langle \hat{a}^\dagger_j \hat{a}_j \rangle } 
    \Bigg) 
    \sim \log \overline{  
    \langle \hat{a}^\dagger_j \hat{a}_j \rangle } \sim \log D_j
    \label{eq:S1thermal}
\end{equation}
where the first approximation holds for large particle numbers.  
If this reasoning were true, then in the non-reciprocal phase, the entropy versus position curve should be peaked at the edges, reflecting the skin-effect-induced localization of the particle number density.  As shown explicitly in Fig.~\ref{fig:thermal_mistake}, this prediction is manifestly incorrect. 
The numerics here matche the analytic arguments presented above:  the true single-site entropy is almost independent of position, and shows no signature of localization.  
As such, simply understanding how average particle number depends on system parameters does not immediately let one understand entanglement properties. The discrepancy between particle number and EE is further explored in App.~\ref{app:sqplth_ansatz}.


\section{Generalized Gibbs ensemble \label{sec:GGE}}

In this last section, we show how the previous results can be extended
to understand the entanglement entropy of small subsystems for size $l$ satisfying $\frac{l}{N}\ll1$. To do this, we make a 
local thermalization hypothesis that the subsystem will be described by a  Generalized Gibbs
ensemble (GGE) state \citep{OriginalGGEpaper,GGERigolVidmar}. The GGE ansatz amounts to the assumption that expectation
value of local observables can be extracted from 
\begin{equation}
\hat\rho_{{\rm GGE}}:=\frac{1}{Z}e^{\sum_{n}\beta_{n}\hat{b}_{n}^{\dagger}\hat{b}_{n}+\gamma_{n}\hat{b}_{\bar{n}}^{\dagger}\hat{b}_{n}^{\dagger}+\gamma_{n}^{*}\hat{b}_{n}\hat{b}_{\bar{n}}}.
\end{equation}
where we recall that the $\hat{b}_n$  modes refer to the standing wave basis, $Z$ is a normalization factor, $\{\beta_{n},\gamma_{n},\gamma_{n}^{*}\}$
are thermodynamic variables fixed by the initial condition and $\bar{n}$
is defined by the relation $\varepsilon_{\bar{n}}=-\varepsilon_{n}$.
Note the contribution of the pairing terms $\hat{b}_{\bar{n}}^{\dagger}\hat{b}_{n}^{\dagger}$,
$\hat{b}_{n}\hat{b}_{\bar{n}}$ that do not appear in $\hat{H}$.
For free bosonic system, this is simply the time average of $\hat {\rho}$
: $\hat\rho_{{\rm GGE}}=\overline{\hat\rho}$. 

In a similar fashion, the GGE ansatz for entanglement is that the EE of a subsystem $A$ can be extracted from $\hat\rho_{{\rm GGE}}$ \citep{ScipostQuasiparticleintegrablesytems},
i.e.~that one has : 
\begin{equation}
S_{A}^{\rm GGE}=\frac{l}{N}\sum_{n=1}^{N}s(\nu_{{\rm GGE}}^{n})
\end{equation}
where $\nu_{{\rm GGE}}^{n}$ are the positive eigenvalues associated to $\sigma_{{\rm GGE}}\Omega$. Given the breaking of translational invariance in the steady state, as exemplified by the strong inhomogeneity in the local occupation number as seen on Fig.~\ref{fig:occ_and_page} \textbf{b}, and the fact that the GGE is agnostic about the cut chosen for $A$, one may expect this approach to fail. On the other hand, the fact that the local correlations are spatially uniform in the tight-binding frame suggests that, for the purpose of computing the EE, the GGE might be enough.

In the non-reciprocal phase, fixing our gauge parameter to $x_0=0$, the positive eigenvalue of (\ref{eq:blockmatrixGGE}) in the continuous limit
is  (see App.~\ref{sec:Equivalence-between-single} for the derivation)
\begin{align}
\nu_{p} & =\sqrt{\cosh^{2}(2r_{0})\left(\left(\frac{\left(p\xi\right)^{2}}{1+\left(p\xi\right)^{2}}\frac{\sinh(L/\xi)}{L/\xi}\right)^{2}-1\right)+1}.\label{eq:nunonrecp-1}
\end{align}
We see that the localization length $\xi$ is the natural scale separating
the long and short wavelength physics. 
The suppression of $\nu_p$ for $p \xi \ll 1$ is a direct consequence of low-momentum standing waves having small wavefunction amplitudes near the system boundary.

In the localized regime $\xi\to0$,
the dominant contribution in the above expression is $\sinh^{2}(L/\xi)$.
Since the entanglement is proportional to the log, we have 
\begin{equation}
s(\nu_{p})\approx\log\nu_{p}\approx L/\xi.
\end{equation}
We see that, because of the exponential scaling of the $\nu_{p}$
with $L$, the different modes' contribution to the entanglement becomes
\emph{independent }of $p$. 

Remarkably, a similar statement is true close to the transition where
$\xi\to\infty$. In this regime, the momentum dependence in (\ref{eq:nunonrecp-1})
cancels out and we are left with 
\begin{align}
\nu_{p} & \approx\cosh(2r_{0})\sqrt{\left(\frac{\sinh(L/\xi)}{L/\xi}\right)^{2}-1}
\end{align}
which is again independent of $p$. 

Similarly, in the reciprocal phase we have
\begin{equation}
\nu_{p}=\sqrt{1+\sinh^{2}\left(2r_{0}\right)\left(1-\frac{32p^{4}\left(1-\cos\left(\varphi L\right)\right)}{\left(\varphi L\left(4p^{2}-\varphi^{2}\right)\right)^{2}}\right)},\label{eq:nurecp-1-1}
\end{equation}
where we defined $\varphi:=\frac{1}{a}(\pi-2\phi)$. For $\varphi$ finite, this quantity becomes $p$ independent
in the large $L$ limit. Close to the critical point, $\varphi\to0$
keeping $\varphi L$ finite while $L\to\infty$ gives 
\begin{equation}
\nu_{p}\approx\sqrt{1+\sinh^{2}\left(2r_{0}\right)\left(1-\frac{2\left(1-\cos\left(\varphi L\right)\right)}{\left(\varphi L\right)^{2}}\right)}\label{eq:nurecp-1-1-1}
\end{equation}
which is also $p$ independent. Thus, we see that in all regimes
of interest, for the purpose of
computing entanglement, the momentum dependence drops out. This in turns implies that the
GGE and the minimal bipartition will match in all the limits mentioned
above and thus, 
\begin{equation}
    S_A^{\rm GGE}=l\overline{S_1}.
\end{equation}
Interestingly, this means that for computing the EE, the local thermalization assumption yields accurate results, despite the fact the system is both strongly inhomogeneous and subject to exponentially large fluctuations. Our interpretation is that fluctuations in local quantities mainly comes from a variation in the squeezing strength which leaves the EE property unchanged. 

\section{Conclusion\label{sec:conclusion}}

Our work demonstrates the existence of an entanglement phase transition in a non-disordered bosonic system undergoing purely unitary evolution. When varying the hopping parameter $g$ below a critical value $\Delta$ the system undergoes a transition from a reciprocal to a non-reciprocal phase, accompanied by a transition from a volume law to a \emph{super-volume} law for the post-quench entanglement entropy of a subsystem. While our system shares many common features with non-Hermitian systems, it {\it does not} involve measurement or post-selection in any way.  Our study suggests that the breaking of reciprocity can be associated with entanglement transitions even in settings where there is no competition between unitary dynamics and measurement-induced non-unitary evolution.  


It is interesting to contrast our results with
the related non-Hermitian fermionic model studied in \citep{Kawabata_PRX_2023_EPT_from_NHSE}, involving two coupled Hatano-Nelson chains.
As discussed, that model exhibits identical spectral properties and NHSE as our system.  
Ref.~\citep{Kawabata_PRX_2023_EPT_from_NHSE} also found an entanglement transition coinciding with the breaking of reciprocity, but unlike us, found that entanglement generation was greatly suppressed in the non-reciprocal phase (yielding only area law behaviour).  In contrast, our non-reciprocal phase exhibits marked directional transport, but no area law entanglement behaviour (and in fact has enhanced entanglement scaling).  This suggests that the  uni-directional quasi-particle picture proposed in \citep{Kawabata_PRX_2023_EPT_from_NHSE} is not applicable to generic entanglement transitions associated with reciprocity breaking.  


It is also interesting to note that in contrast to other studies of bosonic systems, we observe the existence of an EPT despite the absence of measurements \citep{BosonsXiao,MeasurementBosonsRabl} and non-linearities (see e.g.~\cite{ChenFisherLucas_PRRnonlinear}). We also mention that for fermionic systems, entanglement transition were observed for free, unitary, disordered systems; these were directly tied to either Anderson or many-body localization/delocalization transitions, see e.g.~\cite{JiaEEtransitionDisorder,PollmannMooreEEtransitionMBL,GullansHuseEETransitionDisorder2}.
These disorder-driven EPT are also distinct from the phenomenon we describe, as (apart from boundaries) our system is fully translationally invariant.  

While our focus in this work was on post-quench entanglement entropy, it is important to note that the reciprocity-breaking transition in our model can also be characterized with other quantities.  This comprises the spectrally-heralded reciprocal-to-nonreciprocal transition already pointed out in \cite{McDonald_PRX_2018_BKC}. Another observable that shows clear signatures of the transition is the scaling of the total particle number with $N$ in the post-quench state, a quantity which is  \emph{linear} in $\rho$. 
Such signatures of the transition differs markedly from the phenomenology of standard MiPT,  where the transition can be a priori only be characterized using quantities non-linear in $\rho$. 
We stress that the phase transition in the non-Hermitian model of \citep{Kawabata_PRX_2023_EPT_from_NHSE} could also be characterized using a single observable, the total current.  
Returning to our model, we stress that even though the reciprocal and non-reciprocal phases differ strongly in terms of their average density, this does not by itself let one infer the existence of an EPT.  In general, particle number can be made arbitrarily large by means of local squeezing transformations, something that would have no impact on entanglement.    
The fact that average density and entanglement properties can be extremely different is demonstrated explicitly in Fig.~\ref{fig:thermal_mistake} and App.~\ref{app:sqplth_ansatz}, where we observe that the entanglement entropy spatial structure is dramatically different from that of the average particle number. 

The EPT demonstrated in this work is experimentally appealing for several reasons. First, since the model is a non-disordered closed system, post-selection is a complete non-issue. Second, all the studied dynamics are Gaussian, which for bosonic systems are generally considered much more experimentally tractable. Finally, as we showed in Sec.~\ref{sec:SS_analysis}, the entire EPT can be characterized by a single-site covariance matrix. Hence, to detect and characterize the EPT experimentally, one only needs to characterize the correlations of a single site.

In this work, we have demonstrated and characterized an EPT associated with a transition from non-reciprocity to reciprocity in a particular model, namely, the BKC. Future work could investigate the more general relationship between non-reciprocity and entanglement-- in particular, how many of the features of this EPT generalize to other models, and what one can say more generally about the entanglement properties of non-reciprocal models? Finally, we note that while entanglement is a quantum property, one could also investigate a classical version of this model and ask whether the non-reciprocal to reciprocal transition there is also heralded by a transition in correlation measures besides entanglement.  


\begin{acknowledgements}
We thank Vincenzo Alba and Gilles Parez for useful discussions. 
This work was supported by the Air Force Office of Scientific Research under Grant No. FA9550-19-1-0362.  
A. C. also acknowledges support from the Simons Foundation through a Simons Investigator Award (Grant No. 669487, A. C.).

\end{acknowledgements}
\bibliography{references}

\begin{thebibliography}{62}%
\makeatletter
\providecommand \@ifxundefined [1]{%
 \@ifx{#1\undefined}
}%
\providecommand \@ifnum [1]{%
 \ifnum #1\expandafter \@firstoftwo
 \else \expandafter \@secondoftwo
 \fi
}%
\providecommand \@ifx [1]{%
 \ifx #1\expandafter \@firstoftwo
 \else \expandafter \@secondoftwo
 \fi
}%
\providecommand \natexlab [1]{#1}%
\providecommand \enquote  [1]{``#1''}%
\providecommand \bibnamefont  [1]{#1}%
\providecommand \bibfnamefont [1]{#1}%
\providecommand \citenamefont [1]{#1}%
\providecommand \href@noop [0]{\@secondoftwo}%
\providecommand \href [0]{\begingroup \@sanitize@url \@href}%
\providecommand \@href[1]{\@@startlink{#1}\@@href}%
\providecommand \@@href[1]{\endgroup#1\@@endlink}%
\providecommand \@sanitize@url [0]{\catcode `\\12\catcode `\$12\catcode
  `\&12\catcode `\#12\catcode `\^12\catcode `\_12\catcode `\%12\relax}%
\providecommand \@@startlink[1]{}%
\providecommand \@@endlink[0]{}%
\providecommand \url  [0]{\begingroup\@sanitize@url \@url }%
\providecommand \@url [1]{\endgroup\@href {#1}{\urlprefix }}%
\providecommand \urlprefix  [0]{URL }%
\providecommand \Eprint [0]{\href }%
\providecommand \doibase [0]{https://doi.org/}%
\providecommand \selectlanguage [0]{\@gobble}%
\providecommand \bibinfo  [0]{\@secondoftwo}%
\providecommand \bibfield  [0]{\@secondoftwo}%
\providecommand \translation [1]{[#1]}%
\providecommand \BibitemOpen [0]{}%
\providecommand \bibitemStop [0]{}%
\providecommand \bibitemNoStop [0]{.\EOS\space}%
\providecommand \EOS [0]{\spacefactor3000\relax}%
\providecommand \BibitemShut  [1]{\csname bibitem#1\endcsname}%
\let\auto@bib@innerbib\@empty
\bibitem [{\citenamefont {Cao}\ \emph {et~al.}(2019)\citenamefont {Cao},
  \citenamefont {Tilloy},\ and\ \citenamefont {Luca}}]{MiPTCaoTilloyDeLuca}%
  \BibitemOpen
  \bibfield  {author} {\bibinfo {author} {\bibfnamefont {X.}~\bibnamefont
  {Cao}}, \bibinfo {author} {\bibfnamefont {A.}~\bibnamefont {Tilloy}},\ and\
  \bibinfo {author} {\bibfnamefont {A.~D.}\ \bibnamefont {Luca}},\ }\bibfield
  {title} {\bibinfo {title} {{Entanglement in a fermion chain under continuous
  monitoring}},\ }\href {https://doi.org/10.21468/SciPostPhys.7.2.024}
  {\bibfield  {journal} {\bibinfo  {journal} {SciPost Phys.}\ }\textbf
  {\bibinfo {volume} {7}},\ \bibinfo {pages} {024} (\bibinfo {year}
  {2019})}\BibitemShut {NoStop}%
\bibitem [{\citenamefont {Skinner}\ \emph {et~al.}(2019)\citenamefont
  {Skinner}, \citenamefont {Ruhman},\ and\ \citenamefont
  {Nahum}}]{Skinner_PRX_2019_MiPT}%
  \BibitemOpen
  \bibfield  {author} {\bibinfo {author} {\bibfnamefont {B.}~\bibnamefont
  {Skinner}}, \bibinfo {author} {\bibfnamefont {J.}~\bibnamefont {Ruhman}},\
  and\ \bibinfo {author} {\bibfnamefont {A.}~\bibnamefont {Nahum}},\ }\bibfield
   {title} {\bibinfo {title} {Measurement-induced phase transitions in the
  dynamics of entanglement},\ }\href
  {https://doi.org/10.1103/PhysRevX.9.031009} {\bibfield  {journal} {\bibinfo
  {journal} {Phys. Rev. X}\ }\textbf {\bibinfo {volume} {9}},\ \bibinfo {pages}
  {031009} (\bibinfo {year} {2019})}\BibitemShut {NoStop}%
\bibitem [{\citenamefont {Li}\ \emph {et~al.}(2018)\citenamefont {Li},
  \citenamefont {Chen},\ and\ \citenamefont {Fisher}}]{Li_PRB_2018_QZeno}%
  \BibitemOpen
  \bibfield  {author} {\bibinfo {author} {\bibfnamefont {Y.}~\bibnamefont
  {Li}}, \bibinfo {author} {\bibfnamefont {X.}~\bibnamefont {Chen}},\ and\
  \bibinfo {author} {\bibfnamefont {M.~P.~A.}\ \bibnamefont {Fisher}},\
  }\bibfield  {title} {\bibinfo {title} {Quantum zeno effect and the many-body
  entanglement transition},\ }\href
  {https://doi.org/10.1103/PhysRevB.98.205136} {\bibfield  {journal} {\bibinfo
  {journal} {Phys. Rev. B}\ }\textbf {\bibinfo {volume} {98}},\ \bibinfo
  {pages} {205136} (\bibinfo {year} {2018})}\BibitemShut {NoStop}%
\bibitem [{\citenamefont {Gullans}\ \emph {et~al.}(2021)\citenamefont
  {Gullans}, \citenamefont {Krastanov}, \citenamefont {Huse}, \citenamefont
  {Jiang},\ and\ \citenamefont
  {Flammia}}]{Gullans_PRX_2021_quantum_coding_MiPT}%
  \BibitemOpen
  \bibfield  {author} {\bibinfo {author} {\bibfnamefont {M.~J.}\ \bibnamefont
  {Gullans}}, \bibinfo {author} {\bibfnamefont {S.}~\bibnamefont {Krastanov}},
  \bibinfo {author} {\bibfnamefont {D.~A.}\ \bibnamefont {Huse}}, \bibinfo
  {author} {\bibfnamefont {L.}~\bibnamefont {Jiang}},\ and\ \bibinfo {author}
  {\bibfnamefont {S.~T.}\ \bibnamefont {Flammia}},\ }\bibfield  {title}
  {\bibinfo {title} {Quantum coding with low-depth random circuits},\ }\href
  {https://doi.org/10.1103/PhysRevX.11.031066} {\bibfield  {journal} {\bibinfo
  {journal} {Phys. Rev. X}\ }\textbf {\bibinfo {volume} {11}},\ \bibinfo
  {pages} {031066} (\bibinfo {year} {2021})}\BibitemShut {NoStop}%
\bibitem [{\citenamefont {Bao}\ \emph {et~al.}(2020)\citenamefont {Bao},
  \citenamefont {Choi},\ and\ \citenamefont
  {Altman}}]{Bao_PRB_2020_theory_of_MiPT}%
  \BibitemOpen
  \bibfield  {author} {\bibinfo {author} {\bibfnamefont {Y.}~\bibnamefont
  {Bao}}, \bibinfo {author} {\bibfnamefont {S.}~\bibnamefont {Choi}},\ and\
  \bibinfo {author} {\bibfnamefont {E.}~\bibnamefont {Altman}},\ }\bibfield
  {title} {\bibinfo {title} {Theory of the phase transition in random unitary
  circuits with measurements},\ }\href
  {https://doi.org/10.1103/PhysRevB.101.104301} {\bibfield  {journal} {\bibinfo
   {journal} {Phys. Rev. B}\ }\textbf {\bibinfo {volume} {101}},\ \bibinfo
  {pages} {104301} (\bibinfo {year} {2020})}\BibitemShut {NoStop}%
\bibitem [{\citenamefont {Choi}\ \emph {et~al.}(2020)\citenamefont {Choi},
  \citenamefont {Bao}, \citenamefont {Qi},\ and\ \citenamefont
  {Altman}}]{Choi_PRL_2020_Natural_QEC}%
  \BibitemOpen
  \bibfield  {author} {\bibinfo {author} {\bibfnamefont {S.}~\bibnamefont
  {Choi}}, \bibinfo {author} {\bibfnamefont {Y.}~\bibnamefont {Bao}}, \bibinfo
  {author} {\bibfnamefont {X.-L.}\ \bibnamefont {Qi}},\ and\ \bibinfo {author}
  {\bibfnamefont {E.}~\bibnamefont {Altman}},\ }\bibfield  {title} {\bibinfo
  {title} {Quantum error correction in scrambling dynamics and
  measurement-induced phase transition},\ }\href
  {https://doi.org/10.1103/PhysRevLett.125.030505} {\bibfield  {journal}
  {\bibinfo  {journal} {Phys. Rev. Lett.}\ }\textbf {\bibinfo {volume} {125}},\
  \bibinfo {pages} {030505} (\bibinfo {year} {2020})}\BibitemShut {NoStop}%
\bibitem [{\citenamefont {Alberton}\ \emph {et~al.}(2021)\citenamefont
  {Alberton}, \citenamefont {Buchhold},\ and\ \citenamefont
  {Diehl}}]{Alberton_2021}%
  \BibitemOpen
  \bibfield  {author} {\bibinfo {author} {\bibfnamefont {O.}~\bibnamefont
  {Alberton}}, \bibinfo {author} {\bibfnamefont {M.}~\bibnamefont {Buchhold}},\
  and\ \bibinfo {author} {\bibfnamefont {S.}~\bibnamefont {Diehl}},\ }\bibfield
   {title} {\bibinfo {title} {Entanglement transition in a monitored
  free-fermion chain: From extended criticality to area law},\ }\href
  {https://doi.org/10.1103/PhysRevLett.126.170602} {\bibfield  {journal}
  {\bibinfo  {journal} {Phys. Rev. Lett.}\ }\textbf {\bibinfo {volume} {126}},\
  \bibinfo {pages} {170602} (\bibinfo {year} {2021})}\BibitemShut {NoStop}%
\bibitem [{\citenamefont {Potter}\ and\ \citenamefont
  {Vasseur}(2022)}]{Potter_MiPT_Review_2022}%
  \BibitemOpen
  \bibfield  {author} {\bibinfo {author} {\bibfnamefont {A.~C.}\ \bibnamefont
  {Potter}}\ and\ \bibinfo {author} {\bibfnamefont {R.}~\bibnamefont
  {Vasseur}},\ }\bibfield  {title} {\bibinfo {title} {Entanglement dynamics in
  hybrid quantum circuits},\ }in\ \href
  {https://doi.org/10.1007/978-3-031-03998-0_9} {\emph {\bibinfo {booktitle}
  {Quantum Science and Technology}}}\ (\bibinfo  {publisher} {Springer
  International Publishing},\ \bibinfo {year} {2022})\ pp.\ \bibinfo {pages}
  {211--249}\BibitemShut {NoStop}%
\bibitem [{\citenamefont {Fisher}\ \emph {et~al.}(2023)\citenamefont {Fisher},
  \citenamefont {Khemani}, \citenamefont {Nahum},\ and\ \citenamefont
  {Vijay}}]{Fisher_MiPT_Review_2023}%
  \BibitemOpen
  \bibfield  {author} {\bibinfo {author} {\bibfnamefont {M.~P.}\ \bibnamefont
  {Fisher}}, \bibinfo {author} {\bibfnamefont {V.}~\bibnamefont {Khemani}},
  \bibinfo {author} {\bibfnamefont {A.}~\bibnamefont {Nahum}},\ and\ \bibinfo
  {author} {\bibfnamefont {S.}~\bibnamefont {Vijay}},\ }\bibfield  {title}
  {\bibinfo {title} {Random quantum circuits},\ }\href
  {https://doi.org/10.1146/annurev-conmatphys-031720-030658} {\bibfield
  {journal} {\bibinfo  {journal} {Annual Review of Condensed Matter Physics}\
  }\textbf {\bibinfo {volume} {14}},\ \bibinfo {pages} {335} (\bibinfo {year}
  {2023})}\BibitemShut {NoStop}%
\bibitem [{\citenamefont {Buchhold}\ \emph {et~al.}(2021)\citenamefont
  {Buchhold}, \citenamefont {Minoguchi}, \citenamefont {Altland},\ and\
  \citenamefont {Diehl}}]{BuchholdMiPTPRX}%
  \BibitemOpen
  \bibfield  {author} {\bibinfo {author} {\bibfnamefont {M.}~\bibnamefont
  {Buchhold}}, \bibinfo {author} {\bibfnamefont {Y.}~\bibnamefont {Minoguchi}},
  \bibinfo {author} {\bibfnamefont {A.}~\bibnamefont {Altland}},\ and\ \bibinfo
  {author} {\bibfnamefont {S.}~\bibnamefont {Diehl}},\ }\bibfield  {title}
  {\bibinfo {title} {Effective theory for the measurement-induced phase
  transition of dirac fermions},\ }\href
  {https://doi.org/10.1103/PhysRevX.11.041004} {\bibfield  {journal} {\bibinfo
  {journal} {Phys. Rev. X}\ }\textbf {\bibinfo {volume} {11}},\ \bibinfo
  {pages} {041004} (\bibinfo {year} {2021})}\BibitemShut {NoStop}%
\bibitem [{\citenamefont {Turkeshi}\ \emph {et~al.}(2021)\citenamefont
  {Turkeshi}, \citenamefont {Biella}, \citenamefont {Fazio}, \citenamefont
  {Dalmonte},\ and\ \citenamefont {Schir\'o}}]{TurkeshiMiPTinfinitezeroclick}%
  \BibitemOpen
  \bibfield  {author} {\bibinfo {author} {\bibfnamefont {X.}~\bibnamefont
  {Turkeshi}}, \bibinfo {author} {\bibfnamefont {A.}~\bibnamefont {Biella}},
  \bibinfo {author} {\bibfnamefont {R.}~\bibnamefont {Fazio}}, \bibinfo
  {author} {\bibfnamefont {M.}~\bibnamefont {Dalmonte}},\ and\ \bibinfo
  {author} {\bibfnamefont {M.}~\bibnamefont {Schir\'o}},\ }\bibfield  {title}
  {\bibinfo {title} {Measurement-induced entanglement transitions in the
  quantum ising chain: From infinite to zero clicks},\ }\href
  {https://doi.org/10.1103/PhysRevB.103.224210} {\bibfield  {journal} {\bibinfo
   {journal} {Phys. Rev. B}\ }\textbf {\bibinfo {volume} {103}},\ \bibinfo
  {pages} {224210} (\bibinfo {year} {2021})}\BibitemShut {NoStop}%
\bibitem [{\citenamefont {Turkeshi}\ \emph {et~al.}(2020)\citenamefont
  {Turkeshi}, \citenamefont {Fazio},\ and\ \citenamefont
  {Dalmonte}}]{TurkeShiMiPThybrid}%
  \BibitemOpen
  \bibfield  {author} {\bibinfo {author} {\bibfnamefont {X.}~\bibnamefont
  {Turkeshi}}, \bibinfo {author} {\bibfnamefont {R.}~\bibnamefont {Fazio}},\
  and\ \bibinfo {author} {\bibfnamefont {M.}~\bibnamefont {Dalmonte}},\
  }\bibfield  {title} {\bibinfo {title} {Measurement-induced criticality in
  $(2+1)$-dimensional hybrid quantum circuits},\ }\href
  {https://doi.org/10.1103/PhysRevB.102.014315} {\bibfield  {journal} {\bibinfo
   {journal} {Phys. Rev. B}\ }\textbf {\bibinfo {volume} {102}},\ \bibinfo
  {pages} {014315} (\bibinfo {year} {2020})}\BibitemShut {NoStop}%
\bibitem [{\citenamefont {Jin}\ and\ \citenamefont
  {Martin}(2022)}]{JinMartinMiPTclassical}%
  \BibitemOpen
  \bibfield  {author} {\bibinfo {author} {\bibfnamefont {T.}~\bibnamefont
  {Jin}}\ and\ \bibinfo {author} {\bibfnamefont {D.~G.}\ \bibnamefont
  {Martin}},\ }\bibfield  {title} {\bibinfo {title} {Kardar-parisi-zhang
  physics and phase transition in a classical single random walker under
  continuous measurement},\ }\href
  {https://doi.org/10.1103/PhysRevLett.129.260603} {\bibfield  {journal}
  {\bibinfo  {journal} {Phys. Rev. Lett.}\ }\textbf {\bibinfo {volume} {129}},\
  \bibinfo {pages} {260603} (\bibinfo {year} {2022})}\BibitemShut {NoStop}%
\bibitem [{\citenamefont {{Fava}}\ \emph {et~al.}(2023)\citenamefont {{Fava}},
  \citenamefont {{Piroli}}, \citenamefont {{Swann}}, \citenamefont
  {{Bernard}},\ and\ \citenamefont {{Nahum}}}]{NahumNLsM}%
  \BibitemOpen
  \bibfield  {author} {\bibinfo {author} {\bibfnamefont {M.}~\bibnamefont
  {{Fava}}}, \bibinfo {author} {\bibfnamefont {L.}~\bibnamefont {{Piroli}}},
  \bibinfo {author} {\bibfnamefont {T.}~\bibnamefont {{Swann}}}, \bibinfo
  {author} {\bibfnamefont {D.}~\bibnamefont {{Bernard}}},\ and\ \bibinfo
  {author} {\bibfnamefont {A.}~\bibnamefont {{Nahum}}},\ }\bibfield  {title}
  {\bibinfo {title} {{Nonlinear sigma models for monitored dynamics of free
  fermions}},\ }\href {https://doi.org/10.48550/arXiv.2302.12820} {\bibfield
  {journal} {\bibinfo  {journal} {arXiv e-prints}\ ,\ \bibinfo {eid}
  {arXiv:2302.12820}} (\bibinfo {year} {2023})},\ \Eprint
  {https://arxiv.org/abs/2302.12820} {arXiv:2302.12820 [cond-mat.stat-mech]}
  \BibitemShut {NoStop}%
\bibitem [{\citenamefont {{Jian}}\ \emph {et~al.}(2023)\citenamefont {{Jian}},
  \citenamefont {{Shapourian}}, \citenamefont {{Bauer}},\ and\ \citenamefont
  {{Ludwig}}}]{LudwigNLsM}%
  \BibitemOpen
  \bibfield  {author} {\bibinfo {author} {\bibfnamefont {C.-M.}\ \bibnamefont
  {{Jian}}}, \bibinfo {author} {\bibfnamefont {H.}~\bibnamefont
  {{Shapourian}}}, \bibinfo {author} {\bibfnamefont {B.}~\bibnamefont
  {{Bauer}}},\ and\ \bibinfo {author} {\bibfnamefont {A.~W.~W.}\ \bibnamefont
  {{Ludwig}}},\ }\bibfield  {title} {\bibinfo {title} {{Measurement-induced
  entanglement transitions in quantum circuits of non-interacting fermions:
  Born-rule versus forced measurements}},\ }\href
  {https://doi.org/10.48550/arXiv.2302.09094} {\bibfield  {journal} {\bibinfo
  {journal} {arXiv e-prints}\ ,\ \bibinfo {eid} {arXiv:2302.09094}} (\bibinfo
  {year} {2023})},\ \Eprint {https://arxiv.org/abs/2302.09094}
  {arXiv:2302.09094 [cond-mat.stat-mech]} \BibitemShut {NoStop}%
\bibitem [{\citenamefont {{Poboiko}}\ \emph {et~al.}(2023)\citenamefont
  {{Poboiko}}, \citenamefont {{P{\"o}pperl}}, \citenamefont {{Gornyi}},\ and\
  \citenamefont {{Mirlin}}}]{MirlinNLsM}%
  \BibitemOpen
  \bibfield  {author} {\bibinfo {author} {\bibfnamefont {I.}~\bibnamefont
  {{Poboiko}}}, \bibinfo {author} {\bibfnamefont {P.}~\bibnamefont
  {{P{\"o}pperl}}}, \bibinfo {author} {\bibfnamefont {I.~V.}\ \bibnamefont
  {{Gornyi}}},\ and\ \bibinfo {author} {\bibfnamefont {A.~D.}\ \bibnamefont
  {{Mirlin}}},\ }\bibfield  {title} {\bibinfo {title} {{Theory of free fermions
  under random projective measurements}},\ }\href
  {https://doi.org/10.48550/arXiv.2304.03138} {\bibfield  {journal} {\bibinfo
  {journal} {arXiv e-prints}\ ,\ \bibinfo {eid} {arXiv:2304.03138}} (\bibinfo
  {year} {2023})},\ \Eprint {https://arxiv.org/abs/2304.03138}
  {arXiv:2304.03138 [quant-ph]} \BibitemShut {NoStop}%
\bibitem [{\citenamefont {Koh}\ \emph {et~al.}(2023)\citenamefont {Koh},
  \citenamefont {Sun}, \citenamefont {Motta},\ and\ \citenamefont
  {Minnich}}]{Minnich_2023_expt}%
  \BibitemOpen
  \bibfield  {author} {\bibinfo {author} {\bibfnamefont {J.~M.}\ \bibnamefont
  {Koh}}, \bibinfo {author} {\bibfnamefont {S.-N.}\ \bibnamefont {Sun}},
  \bibinfo {author} {\bibfnamefont {M.}~\bibnamefont {Motta}},\ and\ \bibinfo
  {author} {\bibfnamefont {A.~J.}\ \bibnamefont {Minnich}},\ }\bibfield
  {title} {\bibinfo {title} {Measurement-induced entanglement phase transition
  on a superconducting quantum processor with mid-circuit readout},\ }\bibfield
   {journal} {\bibinfo  {journal} {Nat. Phys.}\ }\href
  {https://doi.org/https://doi.org/10.1038/s41567-023-02076-6}
  {https://doi.org/10.1038/s41567-023-02076-6} (\bibinfo {year}
  {2023})\BibitemShut {NoStop}%
\bibitem [{\citenamefont {Gullans}\ and\ \citenamefont
  {Huse}(2020)}]{Gullans_2020}%
  \BibitemOpen
  \bibfield  {author} {\bibinfo {author} {\bibfnamefont {M.~J.}\ \bibnamefont
  {Gullans}}\ and\ \bibinfo {author} {\bibfnamefont {D.~A.}\ \bibnamefont
  {Huse}},\ }\bibfield  {title} {\bibinfo {title} {Scalable probes of
  measurement-induced criticality},\ }\href
  {https://doi.org/10.1103/PhysRevLett.125.070606} {\bibfield  {journal}
  {\bibinfo  {journal} {Phys. Rev. Lett.}\ }\textbf {\bibinfo {volume} {125}},\
  \bibinfo {pages} {070606} (\bibinfo {year} {2020})}\BibitemShut {NoStop}%
\bibitem [{\citenamefont {Ippoliti}\ and\ \citenamefont
  {Khemani}(2021)}]{Ippoliti_2021_xtdual}%
  \BibitemOpen
  \bibfield  {author} {\bibinfo {author} {\bibfnamefont {M.}~\bibnamefont
  {Ippoliti}}\ and\ \bibinfo {author} {\bibfnamefont {V.}~\bibnamefont
  {Khemani}},\ }\bibfield  {title} {\bibinfo {title} {Postselection-free
  entanglement dynamics via spacetime duality},\ }\href
  {https://doi.org/10.1103/PhysRevLett.126.060501} {\bibfield  {journal}
  {\bibinfo  {journal} {Phys. Rev. Lett.}\ }\textbf {\bibinfo {volume} {126}},\
  \bibinfo {pages} {060501} (\bibinfo {year} {2021})}\BibitemShut {NoStop}%
\bibitem [{\citenamefont {Noel}\ \emph {et~al.}(2022)\citenamefont {Noel},
  \citenamefont {Niroula}, \citenamefont {Zhu}, \citenamefont {Risinger},
  \citenamefont {Egan}, \citenamefont {Biswas}, \citenamefont {Cetina},
  \citenamefont {Gorshkov}, \citenamefont {Gullans}, \citenamefont {Huse},\
  and\ \citenamefont {Monroe}}]{NaturePhyMiPTExp}%
  \BibitemOpen
  \bibfield  {author} {\bibinfo {author} {\bibfnamefont {C.}~\bibnamefont
  {Noel}}, \bibinfo {author} {\bibfnamefont {P.}~\bibnamefont {Niroula}},
  \bibinfo {author} {\bibfnamefont {D.}~\bibnamefont {Zhu}}, \bibinfo {author}
  {\bibfnamefont {A.}~\bibnamefont {Risinger}}, \bibinfo {author}
  {\bibfnamefont {L.}~\bibnamefont {Egan}}, \bibinfo {author} {\bibfnamefont
  {D.}~\bibnamefont {Biswas}}, \bibinfo {author} {\bibfnamefont
  {M.}~\bibnamefont {Cetina}}, \bibinfo {author} {\bibfnamefont {A.~V.}\
  \bibnamefont {Gorshkov}}, \bibinfo {author} {\bibfnamefont {M.~J.}\
  \bibnamefont {Gullans}}, \bibinfo {author} {\bibfnamefont {D.~A.}\
  \bibnamefont {Huse}},\ and\ \bibinfo {author} {\bibfnamefont
  {C.}~\bibnamefont {Monroe}},\ }\bibfield  {title} {\bibinfo {title}
  {Measurement-induced quantum phases realized in a trapped-ion quantum
  computer},\ }\href {https://doi.org/10.1038/s41567-022-01619-7} {\bibfield
  {journal} {\bibinfo  {journal} {Nature Physics}\ }\textbf {\bibinfo {volume}
  {18}},\ \bibinfo {pages} {760} (\bibinfo {year} {2022})}\BibitemShut
  {NoStop}%
\bibitem [{\citenamefont {Iadecola}\ \emph {et~al.}(2022)\citenamefont
  {Iadecola}, \citenamefont {Ganeshan}, \citenamefont {Pixley},\ and\
  \citenamefont {Wilson}}]{Iadecola_2022}%
  \BibitemOpen
  \bibfield  {author} {\bibinfo {author} {\bibfnamefont {T.}~\bibnamefont
  {Iadecola}}, \bibinfo {author} {\bibfnamefont {S.}~\bibnamefont {Ganeshan}},
  \bibinfo {author} {\bibfnamefont {J.~H.}\ \bibnamefont {Pixley}},\ and\
  \bibinfo {author} {\bibfnamefont {J.~H.}\ \bibnamefont {Wilson}},\ }\bibfield
   {title} {\bibinfo {title} {Dynamical entanglement transition in the
  probabilistic control of chaos},\ }\href {https://arxiv.org/abs/2207.12415}
  {\bibfield  {journal} {\bibinfo  {journal} {arXiv preprint arXiv:2207.12415}\
  } (\bibinfo {year} {2022})}\BibitemShut {NoStop}%
\bibitem [{\citenamefont {Buchhold}\ \emph {et~al.}(2022)\citenamefont
  {Buchhold}, \citenamefont {M\"{u}ller},\ and\ \citenamefont
  {Diehl}}]{Buchhold_arXiv_2022_preselection}%
  \BibitemOpen
  \bibfield  {author} {\bibinfo {author} {\bibfnamefont {M.}~\bibnamefont
  {Buchhold}}, \bibinfo {author} {\bibfnamefont {T.}~\bibnamefont
  {M\"{u}ller}},\ and\ \bibinfo {author} {\bibfnamefont {S.}~\bibnamefont
  {Diehl}},\ }\href@noop {} {\bibinfo {title} {Revealing measurement-induced
  phase transitions by pre-selection}} (\bibinfo {year} {2022}),\ \Eprint
  {https://arxiv.org/abs/2208.10506} {arXiv:2208.10506 [cond-mat.dis-nn]}
  \BibitemShut {NoStop}%
\bibitem [{\citenamefont {Li}\ \emph {et~al.}(2022)\citenamefont {Li},
  \citenamefont {Zou}, \citenamefont {Glorioso}, \citenamefont {Altman},\ and\
  \citenamefont {Fisher}}]{LiYd_2022}%
  \BibitemOpen
  \bibfield  {author} {\bibinfo {author} {\bibfnamefont {Y.}~\bibnamefont
  {Li}}, \bibinfo {author} {\bibfnamefont {Y.}~\bibnamefont {Zou}}, \bibinfo
  {author} {\bibfnamefont {P.}~\bibnamefont {Glorioso}}, \bibinfo {author}
  {\bibfnamefont {E.}~\bibnamefont {Altman}},\ and\ \bibinfo {author}
  {\bibfnamefont {M.~P.~A.}\ \bibnamefont {Fisher}},\ }\bibfield  {title}
  {\bibinfo {title} {Cross entropy benchmark for measurement-induced phase
  transitions},\ }\href {https://arxiv.org/abs/2209.00609} {\bibfield
  {journal} {\bibinfo  {journal} {arXiv preprint arXiv:2209.00609}\ } (\bibinfo
  {year} {2022})}\BibitemShut {NoStop}%
\bibitem [{\citenamefont {Dehghani}\ \emph {et~al.}(2022)\citenamefont
  {Dehghani}, \citenamefont {Lavasani}, \citenamefont {Hafezi},\ and\
  \citenamefont {Gullans}}]{Dehghani_2022}%
  \BibitemOpen
  \bibfield  {author} {\bibinfo {author} {\bibfnamefont {H.}~\bibnamefont
  {Dehghani}}, \bibinfo {author} {\bibfnamefont {A.}~\bibnamefont {Lavasani}},
  \bibinfo {author} {\bibfnamefont {M.}~\bibnamefont {Hafezi}},\ and\ \bibinfo
  {author} {\bibfnamefont {M.~J.}\ \bibnamefont {Gullans}},\ }\bibfield
  {title} {\bibinfo {title} {Neural-network decoders for measurement induced
  phase transitions},\ }\href {https://arxiv.org/abs/2204.10904} {\bibfield
  {journal} {\bibinfo  {journal} {arXiv preprint arXiv:2204.10904}\ } (\bibinfo
  {year} {2022})}\BibitemShut {NoStop}%
\bibitem [{\citenamefont {Hoke}\ \emph {et~al.}(2023)\citenamefont {Hoke},
  \citenamefont {Ippoliti}, \citenamefont {Abanin}, \citenamefont {Acharya},
  \citenamefont {Ansmann}, \citenamefont {Arute}, \citenamefont {Arya},
  \citenamefont {Asfaw}, \citenamefont {Atalaya}, \citenamefont {Bardin} \emph
  {et~al.}}]{Google_2023_spacetime_duality_expt}%
  \BibitemOpen
  \bibfield  {author} {\bibinfo {author} {\bibfnamefont {J.~C.}\ \bibnamefont
  {Hoke}}, \bibinfo {author} {\bibfnamefont {M.}~\bibnamefont {Ippoliti}},
  \bibinfo {author} {\bibfnamefont {D.}~\bibnamefont {Abanin}}, \bibinfo
  {author} {\bibfnamefont {R.}~\bibnamefont {Acharya}}, \bibinfo {author}
  {\bibfnamefont {M.}~\bibnamefont {Ansmann}}, \bibinfo {author} {\bibfnamefont
  {F.}~\bibnamefont {Arute}}, \bibinfo {author} {\bibfnamefont
  {K.}~\bibnamefont {Arya}}, \bibinfo {author} {\bibfnamefont {A.}~\bibnamefont
  {Asfaw}}, \bibinfo {author} {\bibfnamefont {J.}~\bibnamefont {Atalaya}},
  \bibinfo {author} {\bibfnamefont {J.~C.}\ \bibnamefont {Bardin}}, \emph
  {et~al.},\ }\href@noop {} {\bibinfo {title} {Quantum information phases in
  space-time: measurement-induced entanglement and teleportation on a noisy
  quantum processor}} (\bibinfo {year} {2023}),\ \Eprint
  {https://arxiv.org/abs/2303.04792} {arXiv:2303.04792 [quant-ph]} \BibitemShut
  {NoStop}%
\bibitem [{\citenamefont {Gal}\ \emph {et~al.}(2023)\citenamefont {Gal},
  \citenamefont {Turkeshi},\ and\ \citenamefont
  {Schirò}}]{LeGal_SciPost_2023_NH_MiPT_free_fermions}%
  \BibitemOpen
  \bibfield  {author} {\bibinfo {author} {\bibfnamefont {Y.~L.}\ \bibnamefont
  {Gal}}, \bibinfo {author} {\bibfnamefont {X.}~\bibnamefont {Turkeshi}},\ and\
  \bibinfo {author} {\bibfnamefont {M.}~\bibnamefont {Schirò}},\ }\bibfield
  {title} {\bibinfo {title} {{Volume-to-area law entanglement transition in a
  non-Hermitian free fermionic chain}},\ }\href
  {https://doi.org/10.21468/SciPostPhys.14.5.138} {\bibfield  {journal}
  {\bibinfo  {journal} {SciPost Phys.}\ }\textbf {\bibinfo {volume} {14}},\
  \bibinfo {pages} {138} (\bibinfo {year} {2023})}\BibitemShut {NoStop}%
\bibitem [{\citenamefont {Kawabata}\ \emph {et~al.}(2023)\citenamefont
  {Kawabata}, \citenamefont {Numasawa},\ and\ \citenamefont
  {Ryu}}]{Kawabata_PRX_2023_EPT_from_NHSE}%
  \BibitemOpen
  \bibfield  {author} {\bibinfo {author} {\bibfnamefont {K.}~\bibnamefont
  {Kawabata}}, \bibinfo {author} {\bibfnamefont {T.}~\bibnamefont {Numasawa}},\
  and\ \bibinfo {author} {\bibfnamefont {S.}~\bibnamefont {Ryu}},\ }\bibfield
  {title} {\bibinfo {title} {Entanglement phase transition induced by the
  non-hermitian skin effect},\ }\href
  {https://doi.org/10.1103/PhysRevX.13.021007} {\bibfield  {journal} {\bibinfo
  {journal} {Phys. Rev. X}\ }\textbf {\bibinfo {volume} {13}},\ \bibinfo
  {pages} {021007} (\bibinfo {year} {2023})}\BibitemShut {NoStop}%
\bibitem [{\citenamefont {Gopalakrishnan}\ and\ \citenamefont
  {Gullans}(2021)}]{Gopalakrishnan_PRL_2021_NHEPT}%
  \BibitemOpen
  \bibfield  {author} {\bibinfo {author} {\bibfnamefont {S.}~\bibnamefont
  {Gopalakrishnan}}\ and\ \bibinfo {author} {\bibfnamefont {M.~J.}\
  \bibnamefont {Gullans}},\ }\bibfield  {title} {\bibinfo {title} {Entanglement
  and purification transitions in non-hermitian quantum mechanics},\ }\href
  {https://doi.org/10.1103/PhysRevLett.126.170503} {\bibfield  {journal}
  {\bibinfo  {journal} {Phys. Rev. Lett.}\ }\textbf {\bibinfo {volume} {126}},\
  \bibinfo {pages} {170503} (\bibinfo {year} {2021})}\BibitemShut {NoStop}%
\bibitem [{\citenamefont {Hatano}\ and\ \citenamefont
  {Nelson}(1996)}]{Hatano_1996_HNmodel}%
  \BibitemOpen
  \bibfield  {author} {\bibinfo {author} {\bibfnamefont {N.}~\bibnamefont
  {Hatano}}\ and\ \bibinfo {author} {\bibfnamefont {D.~R.}\ \bibnamefont
  {Nelson}},\ }\bibfield  {title} {\bibinfo {title} {Localization transitions
  in non-hermitian quantum mechanics},\ }\href
  {https://doi.org/10.1103/PhysRevLett.77.570} {\bibfield  {journal} {\bibinfo
  {journal} {Phys. Rev. Lett.}\ }\textbf {\bibinfo {volume} {77}},\ \bibinfo
  {pages} {570} (\bibinfo {year} {1996})}\BibitemShut {NoStop}%
\bibitem [{\citenamefont {Hatano}\ and\ \citenamefont
  {Nelson}(1997)}]{Hatano_1997_vortexpinning}%
  \BibitemOpen
  \bibfield  {author} {\bibinfo {author} {\bibfnamefont {N.}~\bibnamefont
  {Hatano}}\ and\ \bibinfo {author} {\bibfnamefont {D.~R.}\ \bibnamefont
  {Nelson}},\ }\bibfield  {title} {\bibinfo {title} {Vortex pinning and
  non-hermitian quantum mechanics},\ }\href
  {https://doi.org/10.1103/PhysRevB.56.8651} {\bibfield  {journal} {\bibinfo
  {journal} {Phys. Rev. B}\ }\textbf {\bibinfo {volume} {56}},\ \bibinfo
  {pages} {8651} (\bibinfo {year} {1997})}\BibitemShut {NoStop}%
\bibitem [{\citenamefont {Yao}\ \emph {et~al.}(2018)\citenamefont {Yao},
  \citenamefont {Song},\ and\ \citenamefont {Wang}}]{Yao_PRL_2018_NHSE_OG_2}%
  \BibitemOpen
  \bibfield  {author} {\bibinfo {author} {\bibfnamefont {S.}~\bibnamefont
  {Yao}}, \bibinfo {author} {\bibfnamefont {F.}~\bibnamefont {Song}},\ and\
  \bibinfo {author} {\bibfnamefont {Z.}~\bibnamefont {Wang}},\ }\bibfield
  {title} {\bibinfo {title} {Non-hermitian chern bands},\ }\href
  {https://doi.org/10.1103/PhysRevLett.121.136802} {\bibfield  {journal}
  {\bibinfo  {journal} {Phys. Rev. Lett.}\ }\textbf {\bibinfo {volume} {121}},\
  \bibinfo {pages} {136802} (\bibinfo {year} {2018})}\BibitemShut {NoStop}%
\bibitem [{\citenamefont {Martinez~Alvarez}\ \emph {et~al.}(2018)\citenamefont
  {Martinez~Alvarez}, \citenamefont {Barrios~Vargas},\ and\ \citenamefont
  {Foa~Torres}}]{Martinez_PRB_2018_NHSE_OG_3}%
  \BibitemOpen
  \bibfield  {author} {\bibinfo {author} {\bibfnamefont {V.~M.}\ \bibnamefont
  {Martinez~Alvarez}}, \bibinfo {author} {\bibfnamefont {J.~E.}\ \bibnamefont
  {Barrios~Vargas}},\ and\ \bibinfo {author} {\bibfnamefont {L.~E.~F.}\
  \bibnamefont {Foa~Torres}},\ }\bibfield  {title} {\bibinfo {title}
  {Non-hermitian robust edge states in one dimension: Anomalous localization
  and eigenspace condensation at exceptional points},\ }\href
  {https://doi.org/10.1103/PhysRevB.97.121401} {\bibfield  {journal} {\bibinfo
  {journal} {Phys. Rev. B}\ }\textbf {\bibinfo {volume} {97}},\ \bibinfo
  {pages} {121401} (\bibinfo {year} {2018})}\BibitemShut {NoStop}%
\bibitem [{\citenamefont {Kunst}\ \emph {et~al.}(2018)\citenamefont {Kunst},
  \citenamefont {Edvardsson}, \citenamefont {Budich},\ and\ \citenamefont
  {Bergholtz}}]{Kunst_2018_PRL_biorth_correspondence}%
  \BibitemOpen
  \bibfield  {author} {\bibinfo {author} {\bibfnamefont {F.~K.}\ \bibnamefont
  {Kunst}}, \bibinfo {author} {\bibfnamefont {E.}~\bibnamefont {Edvardsson}},
  \bibinfo {author} {\bibfnamefont {J.~C.}\ \bibnamefont {Budich}},\ and\
  \bibinfo {author} {\bibfnamefont {E.~J.}\ \bibnamefont {Bergholtz}},\
  }\bibfield  {title} {\bibinfo {title} {Biorthogonal bulk-boundary
  correspondence in non-hermitian systems},\ }\href
  {https://doi.org/10.1103/PhysRevLett.121.026808} {\bibfield  {journal}
  {\bibinfo  {journal} {Phys. Rev. Lett.}\ }\textbf {\bibinfo {volume} {121}},\
  \bibinfo {pages} {026808} (\bibinfo {year} {2018})}\BibitemShut {NoStop}%
\bibitem [{\citenamefont {McDonald}\ \emph {et~al.}(2018)\citenamefont
  {McDonald}, \citenamefont {Pereg-Barnea},\ and\ \citenamefont
  {Clerk}}]{McDonald_PRX_2018_BKC}%
  \BibitemOpen
  \bibfield  {author} {\bibinfo {author} {\bibfnamefont {A.}~\bibnamefont
  {McDonald}}, \bibinfo {author} {\bibfnamefont {T.}~\bibnamefont
  {Pereg-Barnea}},\ and\ \bibinfo {author} {\bibfnamefont {A.~A.}\ \bibnamefont
  {Clerk}},\ }\bibfield  {title} {\bibinfo {title} {Phase-dependent chiral
  transport and effective non-hermitian dynamics in a bosonic kitaev-majorana
  chain},\ }\href {https://doi.org/10.1103/PhysRevX.8.041031} {\bibfield
  {journal} {\bibinfo  {journal} {Phys. Rev. X}\ }\textbf {\bibinfo {volume}
  {8}},\ \bibinfo {pages} {041031} (\bibinfo {year} {2018})}\BibitemShut
  {NoStop}%
\bibitem [{\citenamefont {Okuma}\ and\ \citenamefont
  {Sato}(2022)}]{Okuma_2022_NH_review}%
  \BibitemOpen
  \bibfield  {author} {\bibinfo {author} {\bibfnamefont {N.}~\bibnamefont
  {Okuma}}\ and\ \bibinfo {author} {\bibfnamefont {M.}~\bibnamefont {Sato}},\
  }\href {https://doi.org/10.48550/ARXIV.2205.10379} {\bibinfo {title}
  {Non-hermitian topological phenomena: A review}} (\bibinfo {year}
  {2022})\BibitemShut {NoStop}%
\bibitem [{\citenamefont {Ashida}\ \emph {et~al.}(2020)\citenamefont {Ashida},
  \citenamefont {Gong},\ and\ \citenamefont {Ueda}}]{Ashida_2020_NH_review}%
  \BibitemOpen
  \bibfield  {author} {\bibinfo {author} {\bibfnamefont {Y.}~\bibnamefont
  {Ashida}}, \bibinfo {author} {\bibfnamefont {Z.}~\bibnamefont {Gong}},\ and\
  \bibinfo {author} {\bibfnamefont {M.}~\bibnamefont {Ueda}},\ }\bibfield
  {title} {\bibinfo {title} {Non-hermitian physics},\ }\href
  {https://doi.org/10.1080/00018732.2021.1876991} {\bibfield  {journal}
  {\bibinfo  {journal} {Advances in Physics}\ }\textbf {\bibinfo {volume}
  {69}},\ \bibinfo {pages} {249} (\bibinfo {year} {2020})},\ \Eprint
  {https://arxiv.org/abs/https://doi.org/10.1080/00018732.2021.1876991}
  {https://doi.org/10.1080/00018732.2021.1876991} \BibitemShut {NoStop}%
\bibitem [{\citenamefont {Bergholtz}\ \emph {et~al.}(2021)\citenamefont
  {Bergholtz}, \citenamefont {Budich},\ and\ \citenamefont
  {Kunst}}]{Bergholtz_2021_RMP_NH_review}%
  \BibitemOpen
  \bibfield  {author} {\bibinfo {author} {\bibfnamefont {E.~J.}\ \bibnamefont
  {Bergholtz}}, \bibinfo {author} {\bibfnamefont {J.~C.}\ \bibnamefont
  {Budich}},\ and\ \bibinfo {author} {\bibfnamefont {F.~K.}\ \bibnamefont
  {Kunst}},\ }\bibfield  {title} {\bibinfo {title} {Exceptional topology of
  non-hermitian systems},\ }\href
  {https://doi.org/10.1103/RevModPhys.93.015005} {\bibfield  {journal}
  {\bibinfo  {journal} {Rev. Mod. Phys.}\ }\textbf {\bibinfo {volume} {93}},\
  \bibinfo {pages} {015005} (\bibinfo {year} {2021})}\BibitemShut {NoStop}%
\bibitem [{\citenamefont {Lin}\ \emph {et~al.}(2023)\citenamefont {Lin},
  \citenamefont {Tai}, \citenamefont {Li},\ and\ \citenamefont
  {Lee}}]{Lin_2023_frontiers_topological_NH_review}%
  \BibitemOpen
  \bibfield  {author} {\bibinfo {author} {\bibfnamefont {R.}~\bibnamefont
  {Lin}}, \bibinfo {author} {\bibfnamefont {T.}~\bibnamefont {Tai}}, \bibinfo
  {author} {\bibfnamefont {L.}~\bibnamefont {Li}},\ and\ \bibinfo {author}
  {\bibfnamefont {C.~H.}\ \bibnamefont {Lee}},\ }\bibfield  {title} {\bibinfo
  {title} {Topological non-hermitian skin effect},\ }\bibfield  {journal}
  {\bibinfo  {journal} {Frontiers of Physics}\ }\textbf {\bibinfo {volume}
  {18}},\ \href {https://doi.org/10.1007/s11467-023-1309-z}
  {10.1007/s11467-023-1309-z} (\bibinfo {year} {2023})\BibitemShut {NoStop}%
\bibitem [{\citenamefont {Wang}\ and\ \citenamefont
  {Clerk}(2019)}]{Wang_PRA_2019_NH_wo_dissipation}%
  \BibitemOpen
  \bibfield  {author} {\bibinfo {author} {\bibfnamefont {Y.-X.}\ \bibnamefont
  {Wang}}\ and\ \bibinfo {author} {\bibfnamefont {A.~A.}\ \bibnamefont
  {Clerk}},\ }\bibfield  {title} {\bibinfo {title} {Non-hermitian dynamics
  without dissipation in quantum systems},\ }\href
  {https://doi.org/10.1103/PhysRevA.99.063834} {\bibfield  {journal} {\bibinfo
  {journal} {Phys. Rev. A}\ }\textbf {\bibinfo {volume} {99}},\ \bibinfo
  {pages} {063834} (\bibinfo {year} {2019})}\BibitemShut {NoStop}%
\bibitem [{\citenamefont {del Pino}\ \emph {et~al.}(2022)\citenamefont {del
  Pino}, \citenamefont {Slim},\ and\ \citenamefont
  {Verhagen}}]{del_Pino_2022_NH_via_sq}%
  \BibitemOpen
  \bibfield  {author} {\bibinfo {author} {\bibfnamefont {J.}~\bibnamefont {del
  Pino}}, \bibinfo {author} {\bibfnamefont {J.~J.}\ \bibnamefont {Slim}},\ and\
  \bibinfo {author} {\bibfnamefont {E.}~\bibnamefont {Verhagen}},\ }\bibfield
  {title} {\bibinfo {title} {Non-hermitian chiral phononics through
  optomechanically induced squeezing},\ }\href
  {https://doi.org/10.1038/s41586-022-04609-0} {\bibfield  {journal} {\bibinfo
  {journal} {Nature}\ }\textbf {\bibinfo {volume} {606}},\ \bibinfo {pages}
  {82} (\bibinfo {year} {2022})}\BibitemShut {NoStop}%
\bibitem [{\citenamefont {Wanjura}\ \emph {et~al.}(2023)\citenamefont
  {Wanjura}, \citenamefont {Slim}, \citenamefont {del Pino}, \citenamefont
  {Brunelli}, \citenamefont {Verhagen},\ and\ \citenamefont
  {Nunnenkamp}}]{Wanjura_2023_quadrature_nonreciprocity}%
  \BibitemOpen
  \bibfield  {author} {\bibinfo {author} {\bibfnamefont {C.~C.}\ \bibnamefont
  {Wanjura}}, \bibinfo {author} {\bibfnamefont {J.~J.}\ \bibnamefont {Slim}},
  \bibinfo {author} {\bibfnamefont {J.}~\bibnamefont {del Pino}}, \bibinfo
  {author} {\bibfnamefont {M.}~\bibnamefont {Brunelli}}, \bibinfo {author}
  {\bibfnamefont {E.}~\bibnamefont {Verhagen}},\ and\ \bibinfo {author}
  {\bibfnamefont {A.}~\bibnamefont {Nunnenkamp}},\ }\bibfield  {title}
  {\bibinfo {title} {Quadrature nonreciprocity in bosonic networks without
  breaking time-reversal symmetry},\ }\bibfield  {journal} {\bibinfo  {journal}
  {Nature Physics}\ }\href {https://doi.org/10.1038/s41567-023-02128-x}
  {10.1038/s41567-023-02128-x} (\bibinfo {year} {2023})\BibitemShut {NoStop}%
\bibitem [{\citenamefont {Zhou}\ and\ \citenamefont
  {Chen}(2021{\natexlab{a}})}]{Zhou_PRB_2021_Gaussian_MiPT_nogo}%
  \BibitemOpen
  \bibfield  {author} {\bibinfo {author} {\bibfnamefont {T.}~\bibnamefont
  {Zhou}}\ and\ \bibinfo {author} {\bibfnamefont {X.}~\bibnamefont {Chen}},\
  }\bibfield  {title} {\bibinfo {title} {Nonunitary entanglement dynamics in
  continuous-variable systems},\ }\href
  {https://doi.org/10.1103/PhysRevB.104.L180301} {\bibfield  {journal}
  {\bibinfo  {journal} {Phys. Rev. B}\ }\textbf {\bibinfo {volume} {104}},\
  \bibinfo {pages} {L180301} (\bibinfo {year}
  {2021}{\natexlab{a}})}\BibitemShut {NoStop}%
\bibitem [{\citenamefont {Minoguchi}\ \emph
  {et~al.}(2022{\natexlab{a}})\citenamefont {Minoguchi}, \citenamefont {Rabl},\
  and\ \citenamefont {Buchhold}}]{Minoguchi_scipost_2022_bosonic_CFT}%
  \BibitemOpen
  \bibfield  {author} {\bibinfo {author} {\bibfnamefont {Y.}~\bibnamefont
  {Minoguchi}}, \bibinfo {author} {\bibfnamefont {P.}~\bibnamefont {Rabl}},\
  and\ \bibinfo {author} {\bibfnamefont {M.}~\bibnamefont {Buchhold}},\
  }\bibfield  {title} {\bibinfo {title} {Continuous gaussian measurements of
  the free boson {CFT}: A model for exactly solvable and detectable
  measurement-induced dynamics},\ }\bibfield  {journal} {\bibinfo  {journal}
  {{SciPost} Physics}\ }\textbf {\bibinfo {volume} {12}},\ \href
  {https://doi.org/10.21468/scipostphys.12.1.009}
  {10.21468/scipostphys.12.1.009} (\bibinfo {year}
  {2022}{\natexlab{a}})\BibitemShut {NoStop}%
\bibitem [{Note1()}]{Note1}%
  \BibitemOpen
  \bibinfo {note} {For $w<\Delta $, both the OBC and the PBC case become
  unstable.}\BibitemShut {Stop}%
\bibitem [{\citenamefont {Kawabata}\ \emph {et~al.}(2019)\citenamefont
  {Kawabata}, \citenamefont {Shiozaki}, \citenamefont {Ueda},\ and\
  \citenamefont {Sato}}]{Kawabata_PRX_2019_NHSE_Topology}%
  \BibitemOpen
  \bibfield  {author} {\bibinfo {author} {\bibfnamefont {K.}~\bibnamefont
  {Kawabata}}, \bibinfo {author} {\bibfnamefont {K.}~\bibnamefont {Shiozaki}},
  \bibinfo {author} {\bibfnamefont {M.}~\bibnamefont {Ueda}},\ and\ \bibinfo
  {author} {\bibfnamefont {M.}~\bibnamefont {Sato}},\ }\bibfield  {title}
  {\bibinfo {title} {Symmetry and topology in non-hermitian physics},\ }\href
  {https://doi.org/10.1103/PhysRevX.9.041015} {\bibfield  {journal} {\bibinfo
  {journal} {Phys. Rev. X}\ }\textbf {\bibinfo {volume} {9}},\ \bibinfo {pages}
  {041015} (\bibinfo {year} {2019})}\BibitemShut {NoStop}%
\bibitem [{\citenamefont {Calabrese}\ and\ \citenamefont
  {Cardy}(2005)}]{CalabreseCardyQuasiparticlepicture_2005}%
  \BibitemOpen
  \bibfield  {author} {\bibinfo {author} {\bibfnamefont {P.}~\bibnamefont
  {Calabrese}}\ and\ \bibinfo {author} {\bibfnamefont {J.}~\bibnamefont
  {Cardy}},\ }\bibfield  {title} {\bibinfo {title} {Evolution of entanglement
  entropy in one-dimensional systems},\ }\href
  {https://doi.org/10.1088/1742-5468/2005/04/P04010} {\bibfield  {journal}
  {\bibinfo  {journal} {Journal of Statistical Mechanics: Theory and
  Experiment}\ }\textbf {\bibinfo {volume} {2005}},\ \bibinfo {pages} {P04010}
  (\bibinfo {year} {2005})}\BibitemShut {NoStop}%
\bibitem [{\citenamefont {Weedbrook}\ \emph {et~al.}(2012)\citenamefont
  {Weedbrook}, \citenamefont {Pirandola}, \citenamefont {Garc\'{\i}a-Patr\'on},
  \citenamefont {Cerf}, \citenamefont {Ralph}, \citenamefont {Shapiro},\ and\
  \citenamefont {Lloyd}}]{RevModPhysGaussianQuantuminformation}%
  \BibitemOpen
  \bibfield  {author} {\bibinfo {author} {\bibfnamefont {C.}~\bibnamefont
  {Weedbrook}}, \bibinfo {author} {\bibfnamefont {S.}~\bibnamefont
  {Pirandola}}, \bibinfo {author} {\bibfnamefont {R.}~\bibnamefont
  {Garc\'{\i}a-Patr\'on}}, \bibinfo {author} {\bibfnamefont {N.~J.}\
  \bibnamefont {Cerf}}, \bibinfo {author} {\bibfnamefont {T.~C.}\ \bibnamefont
  {Ralph}}, \bibinfo {author} {\bibfnamefont {J.~H.}\ \bibnamefont {Shapiro}},\
  and\ \bibinfo {author} {\bibfnamefont {S.}~\bibnamefont {Lloyd}},\ }\bibfield
   {title} {\bibinfo {title} {Gaussian quantum information},\ }\href
  {https://doi.org/10.1103/RevModPhys.84.621} {\bibfield  {journal} {\bibinfo
  {journal} {Rev. Mod. Phys.}\ }\textbf {\bibinfo {volume} {84}},\ \bibinfo
  {pages} {621} (\bibinfo {year} {2012})}\BibitemShut {NoStop}%
\bibitem [{\citenamefont {Hackl}\ and\ \citenamefont
  {Bianchi}(2021)}]{HacklGaussianKahler}%
  \BibitemOpen
  \bibfield  {author} {\bibinfo {author} {\bibfnamefont {L.}~\bibnamefont
  {Hackl}}\ and\ \bibinfo {author} {\bibfnamefont {E.}~\bibnamefont
  {Bianchi}},\ }\bibfield  {title} {\bibinfo {title} {{Bosonic and fermionic
  Gaussian states from Kähler structures}},\ }\href
  {https://doi.org/10.21468/SciPostPhysCore.4.3.025} {\bibfield  {journal}
  {\bibinfo  {journal} {SciPost Phys. Core}\ }\textbf {\bibinfo {volume} {4}},\
  \bibinfo {pages} {025} (\bibinfo {year} {2021})}\BibitemShut {NoStop}%
\bibitem [{\citenamefont {Serafini}(2023)}]{Serafini_2023_QCV_textbook}%
  \BibitemOpen
  \bibfield  {author} {\bibinfo {author} {\bibfnamefont {A.}~\bibnamefont
  {Serafini}},\ }\href@noop {} {\emph {\bibinfo {title} {Quantum continuous
  variables: A Primer of Theoretical Methods}}}\ (\bibinfo  {publisher} {CRC
  Press},\ \bibinfo {year} {2023})\BibitemShut {NoStop}%
\bibitem [{\citenamefont {Page}(1993)}]{PageCurve}%
  \BibitemOpen
  \bibfield  {author} {\bibinfo {author} {\bibfnamefont {D.~N.}\ \bibnamefont
  {Page}},\ }\bibfield  {title} {\bibinfo {title} {Information in black hole
  radiation},\ }\href {https://doi.org/10.1103/PhysRevLett.71.3743} {\bibfield
  {journal} {\bibinfo  {journal} {Phys. Rev. Lett.}\ }\textbf {\bibinfo
  {volume} {71}},\ \bibinfo {pages} {3743} (\bibinfo {year}
  {1993})}\BibitemShut {NoStop}%
\bibitem [{\citenamefont {Fagotti}\ and\ \citenamefont
  {Calabrese}(2008)}]{XYspinChainEEcomputationFagotti}%
  \BibitemOpen
  \bibfield  {author} {\bibinfo {author} {\bibfnamefont {M.}~\bibnamefont
  {Fagotti}}\ and\ \bibinfo {author} {\bibfnamefont {P.}~\bibnamefont
  {Calabrese}},\ }\bibfield  {title} {\bibinfo {title} {Evolution of
  entanglement entropy following a quantum quench: Analytic results for the
  $xy$ chain in a transverse magnetic field},\ }\href
  {https://doi.org/10.1103/PhysRevA.78.010306} {\bibfield  {journal} {\bibinfo
  {journal} {Phys. Rev. A}\ }\textbf {\bibinfo {volume} {78}},\ \bibinfo
  {pages} {010306} (\bibinfo {year} {2008})}\BibitemShut {NoStop}%
\bibitem [{\citenamefont {{Alba}}\ and\ \citenamefont
  {{Calabrese}}(2017)}]{PNASQuasiparticleintegrablesytems}%
  \BibitemOpen
  \bibfield  {author} {\bibinfo {author} {\bibfnamefont {V.}~\bibnamefont
  {{Alba}}}\ and\ \bibinfo {author} {\bibfnamefont {P.}~\bibnamefont
  {{Calabrese}}},\ }\bibfield  {title} {\bibinfo {title} {{Entanglement and
  thermodynamics after a quantum quench in integrable systems}},\ }\href
  {https://doi.org/10.1073/pnas.1703516114} {\bibfield  {journal} {\bibinfo
  {journal} {Proceedings of the National Academy of Science}\ }\textbf
  {\bibinfo {volume} {114}},\ \bibinfo {pages} {7947} (\bibinfo {year}
  {2017})},\ \Eprint {https://arxiv.org/abs/1608.00614} {arXiv:1608.00614
  [cond-mat.str-el]} \BibitemShut {NoStop}%
\bibitem [{\citenamefont {{Parez}}\ \emph {et~al.}(2021)\citenamefont
  {{Parez}}, \citenamefont {{Bonsignori}},\ and\ \citenamefont
  {{Calabrese}}}]{ParezEEExact}%
  \BibitemOpen
  \bibfield  {author} {\bibinfo {author} {\bibfnamefont {G.}~\bibnamefont
  {{Parez}}}, \bibinfo {author} {\bibfnamefont {R.}~\bibnamefont
  {{Bonsignori}}},\ and\ \bibinfo {author} {\bibfnamefont {P.}~\bibnamefont
  {{Calabrese}}},\ }\bibfield  {title} {\bibinfo {title} {{Exact quench
  dynamics of symmetry resolved entanglement in a free fermion chain}},\ }\href
  {https://doi.org/10.1088/1742-5468/ac21d7} {\bibfield  {journal} {\bibinfo
  {journal} {Journal of Statistical Mechanics: Theory and Experiment}\ }\textbf
  {\bibinfo {volume} {2021}},\ \bibinfo {eid} {093102} (\bibinfo {year}
  {2021})},\ \Eprint {https://arxiv.org/abs/2106.13115} {arXiv:2106.13115
  [cond-mat.stat-mech]} \BibitemShut {NoStop}%
\bibitem [{\citenamefont {Rigol}\ \emph {et~al.}(2007)\citenamefont {Rigol},
  \citenamefont {Dunjko}, \citenamefont {Yurovsky},\ and\ \citenamefont
  {Olshanii}}]{OriginalGGEpaper}%
  \BibitemOpen
  \bibfield  {author} {\bibinfo {author} {\bibfnamefont {M.}~\bibnamefont
  {Rigol}}, \bibinfo {author} {\bibfnamefont {V.}~\bibnamefont {Dunjko}},
  \bibinfo {author} {\bibfnamefont {V.}~\bibnamefont {Yurovsky}},\ and\
  \bibinfo {author} {\bibfnamefont {M.}~\bibnamefont {Olshanii}},\ }\bibfield
  {title} {\bibinfo {title} {Relaxation in a completely integrable many-body
  quantum system: An ab initio study of the dynamics of the highly excited
  states of 1d lattice hard-core bosons},\ }\href
  {https://doi.org/10.1103/PhysRevLett.98.050405} {\bibfield  {journal}
  {\bibinfo  {journal} {Phys. Rev. Lett.}\ }\textbf {\bibinfo {volume} {98}},\
  \bibinfo {pages} {050405} (\bibinfo {year} {2007})}\BibitemShut {NoStop}%
\bibitem [{\citenamefont {{Vidmar}}\ and\ \citenamefont
  {{Rigol}}(2016)}]{GGERigolVidmar}%
  \BibitemOpen
  \bibfield  {author} {\bibinfo {author} {\bibfnamefont {L.}~\bibnamefont
  {{Vidmar}}}\ and\ \bibinfo {author} {\bibfnamefont {M.}~\bibnamefont
  {{Rigol}}},\ }\bibfield  {title} {\bibinfo {title} {{Generalized Gibbs
  ensemble in integrable lattice models}},\ }\href
  {https://doi.org/10.1088/1742-5468/2016/06/064007} {\bibfield  {journal}
  {\bibinfo  {journal} {Journal of Statistical Mechanics: Theory and
  Experiment}\ }\textbf {\bibinfo {volume} {6}},\ \bibinfo {pages} {064007}
  (\bibinfo {year} {2016})},\ \Eprint {https://arxiv.org/abs/1604.03990}
  {arXiv:1604.03990 [cond-mat.stat-mech]} \BibitemShut {NoStop}%
\bibitem [{\citenamefont {Alba}\ and\ \citenamefont
  {Calabrese}(2018)}]{ScipostQuasiparticleintegrablesytems}%
  \BibitemOpen
  \bibfield  {author} {\bibinfo {author} {\bibfnamefont {V.}~\bibnamefont
  {Alba}}\ and\ \bibinfo {author} {\bibfnamefont {P.}~\bibnamefont
  {Calabrese}},\ }\bibfield  {title} {\bibinfo {title} {{Entanglement dynamics
  after quantum quenches in generic integrable systems}},\ }\href
  {https://doi.org/10.21468/SciPostPhys.4.3.017} {\bibfield  {journal}
  {\bibinfo  {journal} {SciPost Phys.}\ }\textbf {\bibinfo {volume} {4}},\
  \bibinfo {pages} {017} (\bibinfo {year} {2018})}\BibitemShut {NoStop}%
\bibitem [{\citenamefont {Zhou}\ and\ \citenamefont
  {Chen}(2021{\natexlab{b}})}]{BosonsXiao}%
  \BibitemOpen
  \bibfield  {author} {\bibinfo {author} {\bibfnamefont {T.}~\bibnamefont
  {Zhou}}\ and\ \bibinfo {author} {\bibfnamefont {X.}~\bibnamefont {Chen}},\
  }\bibfield  {title} {\bibinfo {title} {Nonunitary entanglement dynamics in
  continuous-variable systems},\ }\href
  {https://doi.org/10.1103/PhysRevB.104.L180301} {\bibfield  {journal}
  {\bibinfo  {journal} {Phys. Rev. B}\ }\textbf {\bibinfo {volume} {104}},\
  \bibinfo {pages} {L180301} (\bibinfo {year}
  {2021}{\natexlab{b}})}\BibitemShut {NoStop}%
\bibitem [{\citenamefont {Minoguchi}\ \emph
  {et~al.}(2022{\natexlab{b}})\citenamefont {Minoguchi}, \citenamefont {Rabl},\
  and\ \citenamefont {Buchhold}}]{MeasurementBosonsRabl}%
  \BibitemOpen
  \bibfield  {author} {\bibinfo {author} {\bibfnamefont {Y.}~\bibnamefont
  {Minoguchi}}, \bibinfo {author} {\bibfnamefont {P.}~\bibnamefont {Rabl}},\
  and\ \bibinfo {author} {\bibfnamefont {M.}~\bibnamefont {Buchhold}},\
  }\bibfield  {title} {\bibinfo {title} {{Continuous gaussian measurements of
  the free boson CFT: A model for exactly solvable and detectable
  measurement-induced dynamics}},\ }\href
  {https://doi.org/10.21468/SciPostPhys.12.1.009} {\bibfield  {journal}
  {\bibinfo  {journal} {SciPost Phys.}\ }\textbf {\bibinfo {volume} {12}},\
  \bibinfo {pages} {009} (\bibinfo {year} {2022}{\natexlab{b}})}\BibitemShut
  {NoStop}%
\bibitem [{\citenamefont {Chen}\ \emph {et~al.}(2020)\citenamefont {Chen},
  \citenamefont {Li}, \citenamefont {Fisher},\ and\ \citenamefont
  {Lucas}}]{ChenFisherLucas_PRRnonlinear}%
  \BibitemOpen
  \bibfield  {author} {\bibinfo {author} {\bibfnamefont {X.}~\bibnamefont
  {Chen}}, \bibinfo {author} {\bibfnamefont {Y.}~\bibnamefont {Li}}, \bibinfo
  {author} {\bibfnamefont {M.~P.~A.}\ \bibnamefont {Fisher}},\ and\ \bibinfo
  {author} {\bibfnamefont {A.}~\bibnamefont {Lucas}},\ }\bibfield  {title}
  {\bibinfo {title} {Emergent conformal symmetry in nonunitary random dynamics
  of free fermions},\ }\href {https://doi.org/10.1103/PhysRevResearch.2.033017}
  {\bibfield  {journal} {\bibinfo  {journal} {Phys. Rev. Res.}\ }\textbf
  {\bibinfo {volume} {2}},\ \bibinfo {pages} {033017} (\bibinfo {year}
  {2020})}\BibitemShut {NoStop}%
\bibitem [{\citenamefont {Jia}\ \emph {et~al.}(2008)\citenamefont {Jia},
  \citenamefont {Subramaniam}, \citenamefont {Gruzberg},\ and\ \citenamefont
  {Chakravarty}}]{JiaEEtransitionDisorder}%
  \BibitemOpen
  \bibfield  {author} {\bibinfo {author} {\bibfnamefont {X.}~\bibnamefont
  {Jia}}, \bibinfo {author} {\bibfnamefont {A.~R.}\ \bibnamefont
  {Subramaniam}}, \bibinfo {author} {\bibfnamefont {I.~A.}\ \bibnamefont
  {Gruzberg}},\ and\ \bibinfo {author} {\bibfnamefont {S.}~\bibnamefont
  {Chakravarty}},\ }\bibfield  {title} {\bibinfo {title} {Entanglement entropy
  and multifractality at localization transitions},\ }\href
  {https://doi.org/10.1103/PhysRevB.77.014208} {\bibfield  {journal} {\bibinfo
  {journal} {Phys. Rev. B}\ }\textbf {\bibinfo {volume} {77}},\ \bibinfo
  {pages} {014208} (\bibinfo {year} {2008})}\BibitemShut {NoStop}%
\bibitem [{\citenamefont {Bardarson}\ \emph {et~al.}(2012)\citenamefont
  {Bardarson}, \citenamefont {Pollmann},\ and\ \citenamefont
  {Moore}}]{PollmannMooreEEtransitionMBL}%
  \BibitemOpen
  \bibfield  {author} {\bibinfo {author} {\bibfnamefont {J.~H.}\ \bibnamefont
  {Bardarson}}, \bibinfo {author} {\bibfnamefont {F.}~\bibnamefont
  {Pollmann}},\ and\ \bibinfo {author} {\bibfnamefont {J.~E.}\ \bibnamefont
  {Moore}},\ }\bibfield  {title} {\bibinfo {title} {Unbounded growth of
  entanglement in models of many-body localization},\ }\href
  {https://doi.org/10.1103/PhysRevLett.109.017202} {\bibfield  {journal}
  {\bibinfo  {journal} {Phys. Rev. Lett.}\ }\textbf {\bibinfo {volume} {109}},\
  \bibinfo {pages} {017202} (\bibinfo {year} {2012})}\BibitemShut {NoStop}%
\bibitem [{\citenamefont {Gullans}\ and\ \citenamefont
  {Huse}(2019)}]{GullansHuseEETransitionDisorder2}%
  \BibitemOpen
  \bibfield  {author} {\bibinfo {author} {\bibfnamefont {M.~J.}\ \bibnamefont
  {Gullans}}\ and\ \bibinfo {author} {\bibfnamefont {D.~A.}\ \bibnamefont
  {Huse}},\ }\bibfield  {title} {\bibinfo {title} {Localization as an
  entanglement phase transition in boundary-driven anderson models},\ }\href
  {https://doi.org/10.1103/PhysRevLett.123.110601} {\bibfield  {journal}
  {\bibinfo  {journal} {Phys. Rev. Lett.}\ }\textbf {\bibinfo {volume} {123}},\
  \bibinfo {pages} {110601} (\bibinfo {year} {2019})}\BibitemShut {NoStop}%
\end{thebibliography}%

\begin{appendix}

\section{Details on the numerics \label{app:numerics}}

We are interested in the long-time average entanglement that results from the quench dynamics described in the main text, and hence the quantity of interest is \begin{equation}
    \overline{S_t} = \lim_{T \rightarrow \infty} \frac{1}{T} \int_0^T S_t dt,
\end{equation}
which is the EE for some subsystem of our $1D$ BKC lattice for some fixed set of parameters.  Without loss of generality, we will fix $w = 1, \Delta = 0.25$, and vary $g, N$, where $\Delta$ and $g$ will be written in units of $w$. We will estimate $\overline{S_t}$ by numerically calculating $S_t$ for some discrete set of times and then taking the mean. Since we want the EE in the quasi-steady state, we only need to perform this calculation up to some finite large $T$ for which the estimate of $\overline{S_t}$  converges to some desired accuracy. Since the evolution of $S_t$ is deterministic, we can set the accuracy to any level we want. 

In general, one might not expect to be able to simulate arbitrarily large times accurately for non-reciprocal systems, due to the issue of numerical ill-conditioning. Fortunately, we can avoid this by performing the simulations in the squeezing frame defined by Eq.~(\eqref{eq:diag_transform}), something that is possible whenever $g \neq \Delta$. For $g = \Delta$, the squeezing frame is not well-defined. In that case, we simply performed the simulations in the lab frame, and found it to be stable for all chosen parameters in this work.

The following are additional important points about the numerical approach used to calculate the entanglement entropy.  
\begin{enumerate}
    \item The value of the EE at initial small times $t$ are in general not representative of the quasi-steady state of interest. While they get averaged away at long times, including these points slow down the convergence of our calculation. We thus pick an initial time $T_{\rm min}$ at which $S_t$ has approximately relaxed to its quasi-steady state value, and only use $t \geq T_{\rm min}$ to calculate the needed average. In the tight-binding frame, we can read off the group velocity as $J$ and hence we expect the system to relax with time scale $O(N/J)$. For convenience, we pick $T_{\rm min} = 10 N / J$, where $10$ is a reasonably large prefactor.
    \item Next, we pick an initial set of times $\{ T_1, ..., T_{1000} \}$. We do so by picking a uniformly spaced set of times with $T_{1} = T_{\rm min}, T_{k + 1} - T_k = \delta T$. To aid convergence, we want $\delta T$ to be relatively large compared to the time-scale of oscillations in the quasi-steady state, which we generally expect to occur on the time scale $O(1/J)$. Again, we pick $10$ as an arbitrary reasonably large prefactor and set $\delta T = 10/J$.
    \item We numerically calculate the values $\{S_{T_1}, ..., S_{T_{1000}}\}$ and estimate $\widetilde{\overline{S}_t} = \frac{1}{1000} \sum_{i} S_{t_i}$. We use $\widetilde{\varepsilon} = s/\sqrt{N}$, where $s$ is the standard deviation of the set $\{S_{T_1}, ..., S_{T_{1000}}\}$, to estimate the accuracy of $\widetilde{\overline{S}_t}$. 
    \item We want the relative error (compared to the mean) to be small, so we pick an arbitrary harsh convergence criterion $\widetilde{\varepsilon}/\widetilde{\overline{S}_t} < 0.001$. If this is satisfied, we are done. Otherwise, we repeat the procedure for another $500$ time-steps $\{ T_{1001}, ..., T_{1500}\}$ selected in the same way, and check the convergence criterion, repeating until it is satisfied.
\end{enumerate}

Finally, we numerically verified that this simulation is insensitive to the exact values of each of the prefactors stated above. For the convergence threshold $\widetilde{\varepsilon}/\widetilde{\overline{S}_t} < 0.001$ that we have picked, the error bars are not visible on the plots, and we have chosen to omit them. The simulations in Figs.~\ref{fig:BKC_schematic}, \ref{fig:EE_numerics} are performed for $g = 0, 0.2, 0.24, 0.245, 0.249, 0.25, 0.251, 0.255, 0.26$ and $N = 16, 32, 48, 64, 96, 128$.

\section{Full time evolution and fluctuations of EE in the non-reciprocal phase \label{app:fluctuations}}

In this appendix, we numerically study the fluctuations in the EE in the quasi-steady state of the non-reciprocal phase, in order to demonstrate that the mean EE is a meaningful characterization of the quasi-steady state. In other words, fluctuations in EE over time in the long-time regime are sufficiently small or comparable to the mean value. Surprisingly, this is true even though it is not the case for any of the other quantities of interest, such as $\nu_t, \nu_t^2, \overline{\langle \hat{d}^{\dag} \hat{d} \rangle}$, where fluctuations can significantly larger than the mean value. 

\begin{figure*}[tp]
\includegraphics[width=2\columnwidth]{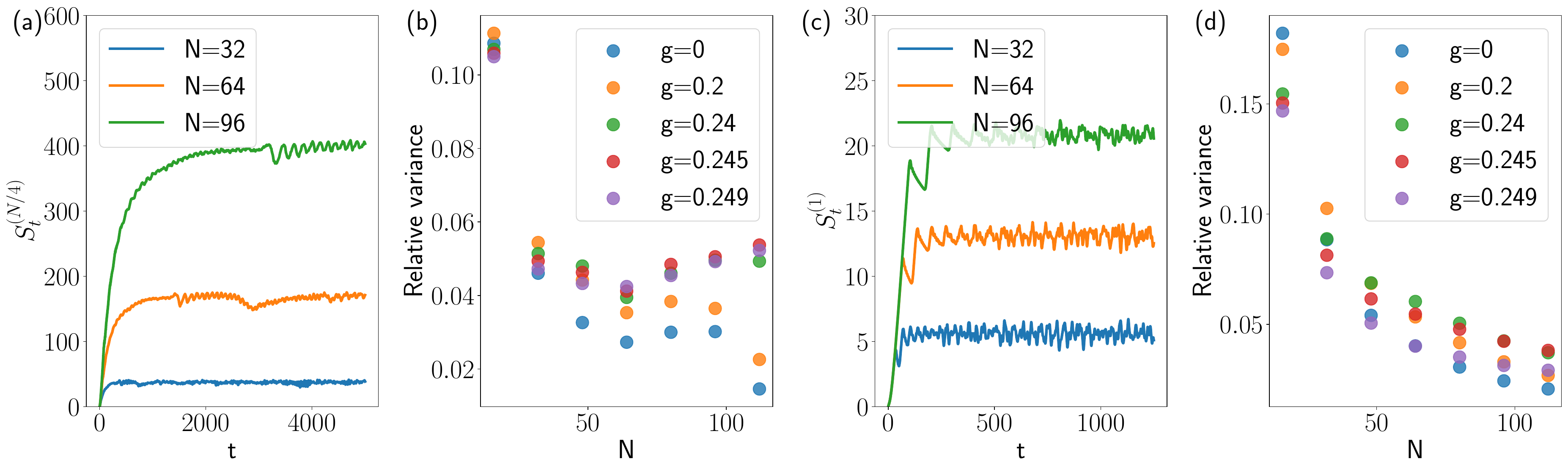}\label{fig:EE_mean_characterization}
\caption{(a) Full time evolution of the EE of the $(N/4:3N/4)$ biparition for a representative time period for $w =1 , \Delta = 0.25, g = 0$. We observe that visually, the long-time EE appears to oscillate around some mean value, with oscillations small compared to the mean value. (b) Plot of the variance defined in Eq.~(\ref{app_eq:S_t_var}), calculated over the period of time required for the mean value of the EE to converge according to the criterion outlined in App.~\ref{app:numerics}. (c) Same as (a) but for the minimal bipartition $(1:N-1)$. (d) Same as (b) but for the minimal bipartition $(1:N-1)$.}
 
\end{figure*}

We expect fluctuations to be largest when non-reciprocity is maximum. In terms of the numerics provided in the main text, this is when $g = 0, \Delta = 0.25, w = 1$. Fig.~\ref{fig:EE_mean_characterization}a, c plot the full time evolution of the EE for $N = 16, 32, 48, 64$ for the $N/4$ bipartition and the minimal bipartition. We observe, at least visually, that the EE does indeed reach a quasi-steady state, with fluctuations over time much smaller than the average value. We can quantify the size of the fluctuations compared to the mean value using the quantity 
\begin{equation}\label{app_eq:S_t_var}
    \frac{\sqrt{\overline{(S_t - \overline{S_t})^2}}}{\overline{S_t}},
\end{equation}
where we calculate the time-averaged quantities using the methods outlined in App.~\ref{app:numerics}. These quantities are plotted in Fig.~\ref{fig:EE_mean_characterization}b, d for values of $N$ up to $N=64$ and $g = 0, 0.2, 0.24, 0.245, 0.249$, and observe that the above quantity generally decreases with $N$, giving a value $ \sim 10^{-2} - 10^{-1}$  across all $N$ considered. We comment that this property does not hold for other quantities, such as $\nu_t$ or particle density. Furthermore, note that we don't require this quantity to go to $0$ -- simply that it is reasonably small enough that the time-averaged EE provides a good description of the quasi-steady state.

\section{Average correlations \label{App : Average correlations}}

In this appendix we derive explicitly the time-averaged covariance
$\overline{\sigma}$ both in the reciprocal and non-reciprocal phases.
In particular, we will show that in the squeezed frame, $\overline{\sigma}$
is \emph{independent }of the position in the lattice. 

In both the reciprocal and non-reciprocal phases, the Hamiltonian
can be brought under the diagonal form $\hat{H}=\sum_{n}\varepsilon_{n}\hat{b}_{n}^{\dagger}\hat{b}_{n}$
with $\varepsilon_{n}=-2\sqrt{w^{2}+g^{2}-\Delta^{2}}\cos\left(\frac{\pi n}{N+1}\right)$.

The time average for the correlations in position basis is particularly
simple and can be written as the selection rules 
\begin{align}
\overline{\langle\hat{b}_{m}^{\dagger}\hat{b}_{n}\rangle} & =\delta_{m,n}\langle\hat{b}_{n}^{\dagger}\hat{b}_{n}\rangle_{t=0},\\
\overline{\langle\hat{b}_{m}\hat{b}_{n}\rangle} & =\delta_{m,\bar{n}}\langle\hat{b}_{\bar{n}}\hat{b}_{n}\rangle_{t=0},
\end{align}
where we introduced $\bar{n}:=N+1-n$. Hence the average state of
the system is entirely determined by the values of these correlations
at $t=0$. 

\subsection{Non-reciprocal phase}

In the non-reciprocal phase, the eigenmodes are given by

\begin{align}
 & \hat{b}_{n}=\sqrt{\frac{2}{N+1}}\sum_{j=1}^{N}e^{-i\frac{\pi j}{2}}\sin\left(\frac{\pi jn}{N+1}\right)\label{eq:changeofbasisnonrecp}\\
 & \bigg(\left(\cosh\left(r(j-j_{0})\right)\cosh r_0+i\sinh\left(r(j-j_{0})\right)\sinh r_0\right)\hat{a}_{j}\nonumber \\
 & +\left(i\cosh\left(r(j-j_{0})\right)\sinh r_0-\sinh\left(r(j-j_{0})\right)\cosh r_0\right)\hat{a}_{j}^{\dagger}\bigg).\nonumber 
\end{align}
where $j_{0}$ is an arbitrary ``gauge factor'' and $\tanh\left(2 r_0\right)=\frac{g}{\Delta}$,
$e^{2r}=\frac{w+\sqrt{\Delta^{2}-g^{2}}}{w-\sqrt{\Delta^{2}-g^{2}}}$.
The conserved correlations are fixed by the vacuum initial state 
\begin{align}\label{app_eq:initial_TB_real_space_corr}
\langle\hat{b}_{n}^{\dagger}\hat{b}_{n}\rangle_{t=0}= & v(n,r)\cosh(2 r_0)-\frac{1}{2},\\
\langle\hat{b}_{\bar{n}}\hat{b}_{n}\rangle_{t=0}= & w(n,r)\cosh(2 r_0)-\frac{i}{2}\sinh(2 r_0),
\end{align}
with 
\begin{align}
v(n,r) & :=\frac{1}{N+1}\sum_{j}\cosh\left(2r(j-j_{0})\right)\sin^{2}\left(\frac{\pi nj}{N+1}\right),\\
w(n,r) & :=\frac{1}{N+1}\sum_{j}\sinh\left(2r(j-j_{0})\right)\sin^{2}\left(\frac{\pi nj}{N+1}\right).
\end{align}
These expressions simplify in the continuous limit defined as follows.
Let $a$ be the lattice spacing. We consider the limit $N\to\infty$,
$a\to0$, while keeping fixed the dimensionful quantities $\xi:=\frac{a}{r}$,
$L:=a(N+1)$, $x:=aj$ and $p:=\frac{\pi n}{a(N+1)}$. To simplify the expressions, we fix the gauge parameter $j_0=0$. This leads
to 
\begin{align}
v(p,\zeta) & =\frac{\xi p^{2}}{2L\left(\xi^{-2}+p^{2}\right)}\frac{\sinh(2L/\xi)}{2},\\
w(p,\zeta) & =\frac{\xi p^{2}}{2L\left(\xi^{-2}+p^{2}\right)}\frac{\cosh{(2L/\xi)}-1}{2}.
\end{align}

\paragraph{Local correlations in the tight-binding basis}

In this part we derive the average local on-site correlations in a
given spatial frame. As discussed in the main text, those are the
quantities necessary to characterize entanglement in the minimal bipartition
protocol. The spatial frame where the correlations appear in their
simplest form is the tight-binding frame with operators $\{\hat{d}_{j}^{\dagger},\hat{d}_{j}\}$.
The correlations are related to the one in the diagonal basis by a
simple OBC Fourier transform : 
\begin{align}
\overline{\langle\hat{d}_{j}^{\dagger}\hat{d}_{j}\rangle} & =\frac{2}{N+1}\sum_{n=1}^{N}\sin^{2}\left(\frac{\pi nj}{N+1}\right)\overline{\langle\hat{b}_{n}^{\dagger}\hat{b}_{n}\rangle},\\
\overline{\langle\hat{d}_{j}\hat{d}_{j}\rangle} & =\frac{2}{N+1}(-1)^{j+1}\sum_{n=1}^{N}\sin^{2}\left(\frac{\pi nj}{N+1}\right)\overline{\langle\hat{b}_{\bar{n}}\hat{b}_{n}\rangle}.
\end{align}
We will make use of the identity
\begin{align}
 & \frac{2}{N+1}\sum_{n=1}^{L}\sin^{2}\left(\frac{\pi nj}{N+1}\right)\sin^{2}\left(\frac{\pi nl}{N+1}\right)\\
 & =\frac{1}{2}+\frac{1}{4}\left(\delta_{j,l}+\delta_{j,N+1-l}\right).
\end{align}
Inserting this identity in the previous relations leads to 
\begin{align}
 & \overline{\langle\hat{d}_{j}^{\dagger}\hat{d}_{j}\rangle}=-\frac{1}{2}\nonumber \\
 & +\frac{1}{N+1}\frac{\cosh(2 r_0)}{2}\bigg(\frac{\sinh(rN)\cosh(r(N+1-2j_{0}))}{\sinh(r)}\nonumber \\
 & +\frac{1}{2}\left(\cosh\left(2r(j-j_{0})\right)+\cosh\left(2r(N+1-j-j_{0})\right)\right)\bigg)
\end{align}
where we recall that $j_{0}$ is an arbitrary ``gauge factor''.
Once again, this expression simplifies in the continuous limit: 
\begin{align}
\overline{\langle\hat{d}_{x}^{\dagger}\hat{d}_{x}\rangle} & =\cosh(2 r_0)\frac{\sinh(2(L-x_{0})/\xi)+\sinh(2 x_{0}/\xi)}{4L/\xi}-\frac{1}{2},
\end{align}
Note that the $j$ dependent term is no longer here in the continuous
limit description. Finally, we can fix the gauge parameter $x_{0}=L/2$ to
simplify these expressions: 
\begin{align}
\overline{\langle\hat{d}_{x}^{\dagger}\hat{d}_{x}\rangle} & =\cosh(2 r_0)\frac{\sinh(L/\xi)}{2L/\xi}-\frac{1}{2}.
\end{align}
For the local pair annihilation correlation, one obtains 

\begin{align}
 & \overline{\langle\hat{d}_{j}\hat{d}_{j}\rangle}=\frac{(-1)^{j+1}}{2}\bigg(-i\sinh(2 r_0)\nonumber\\
 & \frac{1}{N+1}\cosh(2 r_0)\bigg(\frac{\sinh(rN)\sinh(r(N+1-2j_{0}))}{\sinh(r)}\nonumber\\
 & +\frac{1}{2}\left(\sinh\left(2r(j-j_{0})\right)+\sinh\left(2r(N+1-j-j_{0})\right)\right)\bigg)\bigg).
\end{align}
Once again, taking the continuous limit and choosing $x_{0}=L/2$,
one gets 
\begin{align}
\overline{\langle\hat{d}_{x}\hat{d}_{x}\rangle} & =(-1)^{\frac{x}{a}}\frac{i}{2}\sinh(2 r_0).
\end{align}
We thus see that, in the continuous limit defined above, the norms of both correlations
are independent of $x$ in the tight-binding frame. For the minimal
bipartition, this means that the value of the entanglement entropy
will be the same, up to finite size corrections, for all the sites. 

\subsection{Reciprocal phase}

In the reciprocal phase, the eigenoperators are 
\begin{align}
\hat{b}_{n}= & \sqrt{\frac{2}{N+1}}\sum_{j}\sin\left(\frac{\pi nj}{N+1}\right)\nonumber \\
 & e^{-i\phi j}\left(\cosh r_0\hat{a}_{j}+i\sinh r_0\hat{a}_{j}^{\dagger}\right),
\end{align}
with $\tanh\left(2 r_0\right)=\frac{g}{\Delta}$ and $\phi 
= \arctan {(} w/
\sqrt{g^{2}-\Delta^{2}} )$.
The conserved correlations in this case are given by 
\begin{align}
 & \langle\hat{b}_{n}^{\dagger}\hat{b}_{n}\rangle_{t=0}=\overline{\langle\hat{b}_{n}^{\dagger}\hat{b}_{n}\rangle}=\frac{1}{2}\left(\cosh\left(2 r_0\right)-1\right),\\
 & \langle\hat{b}_{\bar{n}}\hat{b}_{n}\rangle_{t=0}=\overline{\langle\hat{b}_{\bar{n}}\hat{b}_{n}\rangle}=\\
 & -i\frac{\sinh\left(2 r_0\right)}{2}\frac{2}{N+1}\sum_{j=1}^{N}\sin^{2}\left(\frac{\pi nj}{N+1}\right)e^{-2i(\phi-\frac{\pi}{2})j}.\nonumber 
\end{align}
Performing the sum for $\overline{\langle\hat{b}_{\bar{n}}\hat{b}_{n}\rangle}$
in the continuous limit leads to 
\begin{equation}
\overline{\langle\hat{b}_{\bar{p}}\hat{b}_{p}\rangle}=\frac{1}{L}\sinh\left(2 r_0\right)\frac{2p^{2}\left(1-e^{i\varphi L}\right)}{\varphi\left(4p^{2}-\varphi^{2}\right)}, 
\end{equation}
where we defined $\bar{p}:=\frac{\pi}{a}-p$ and $\varphi:=\frac{1}{a}(\pi-2\phi)$. 

\paragraph{Local correlations in the tight-binding basis}

As for the non-reciprocal case, the local on-site correlations take
their simplest form in the tight-binding frame with operators $\{\hat{d}_{j}^{\dagger},\hat{d}_{j}\}$.
Performing the inverse Fourier transform leads in this case to: 
\begin{align}
 & \overline{\langle\hat{d}_{j}^{\dagger}\hat{d}_{j}\rangle}=\frac{1}{2}\left(\cosh\left(2 r_0\right)-1\right),\\
 & \overline{\langle\hat{d}_{j}\hat{d}_{j}\rangle}=\\
 & i\frac{(-1)^{j}\sinh\left(2 r_0\right)}{2(N+1)}\left(\frac{e^{i\theta j}+e^{i\theta(N+1-j)}}{2}-\frac{1-e^{i\theta N}}{1-e^{-i\theta}}\right).\nonumber 
\end{align}
Defining $\varphi:=\frac{\theta}{a}$, the last expression simplifies
once again in the continuous limit: 
\begin{equation}
\overline{\langle\hat{d}_{x}\hat{d}_{x}\rangle}=(-1)^{x/a}\frac{\sinh\left(2 r_0\right)}{2}\frac{e^{i\varphi L}-1}{\varphi L}.
\end{equation}

\section{Computation of entanglement entropy \label{sec:Equivalence-between-single}}

In this appendix, we compute the EE of a subsystem of size $l$ in
the limit $l/L\ll1$ using the GGE. We begin by showing that, for
our model, this is equivalent to the minimal bipartition approach,
both in the reciprocal and non-reciprocal phase.

\subsection{Equivalence between GGE ansatz and minimal bipartition}

Recall that in the GGE approach, the stationary entanglement of a
subsystem $A$ of size $l$ is simply assumed to be directly proportional
to the total EE of the total system, with the proportionality coefficient
fixed by $l$, 
\begin{equation}
S_{A}=\frac{l}{N}\sum_{n}s(\nu_{n}).
\end{equation}
with $s(x):=\left(\frac{x+1}{2}\right)\ln\left(\frac{x+1}{2}\right)-\left(\frac{x-1}{2}\right)\ln\left(\frac{x-1}{2}\right)$
and $\nu_{n}$ the positive eigenvalue associated to the $2\times2$
block matrix
\begin{equation}
\begin{pmatrix}2\overline{\langle\hat{b}_{n}^{\dagger}\hat{b}_{n}\rangle}+1 & -2\overline{\langle\hat{b}_{\bar{n}}\hat{b}_{n}\rangle}\\
2\overline{\langle\hat{b}_{\bar{n}}\hat{b}_{n}\rangle}^{*} & -2\overline{\langle\hat{b}_{n}^{\dagger}\hat{b}_{n}\rangle}-1
\end{pmatrix}.\label{eq:blockmatrixGGE}
\end{equation}

\paragraph{Non-reciprocal phase}

Recall the expressions for the correlations in the continuous limit (Recall we fixed $x_0=0$ in this case)
:
\begin{align}
\overline{\langle\hat{b}_{p}^{\dagger}\hat{b}_{p}\rangle}= & \frac{ (p\xi)^{2}}{2\left(1+(p\xi)^{2}\right)}\frac{\sinh(2L/\xi)}{2L/\xi}\cosh(2 r_0)-\frac{1}{2},\\
\overline{\langle\hat{b}_{\bar{p}}\hat{b}_{p}\rangle}= & \frac{(p\xi )^{2}}{1+(p\xi)^{2}}\frac{\sinh^{2}(L/\xi)}{2L/\xi}\cosh(2 r_0)-\frac{i}{2}\sinh(2 r_0)
\end{align}
with $\frac{\pi}{L}\leq p\leq \frac{\pi}{a}$. The positive eigenvalue of (\ref{eq:blockmatrixGGE}) is 
\begin{align}
\nu_{p} & =\sqrt{\cosh^{2}(2 r_0)\left(\left(\frac{\xi p^{2}}{L\left(\xi^{-2}+p^{2}\right)}\right)^{2}\sinh^{2}(L/\xi)-1\right)+1}.\label{eq:nunonrecp}
\end{align}
Away from the critical point, $\zeta$ is finite. Taking the large
system size limit then leads to : 
\begin{align}
\nu_{p} & \approx\frac{\xi p^{2}}{L\left(\xi^{-2}+p^{2}\right)}\cosh(2 r_0)\sinh(L/\xi).
\end{align}
Importantly 
\begin{equation}
s(\nu_{p})\approx\ln\nu_{p}\approx L/\xi,
\end{equation}
to leading order. Thus we see that to leading order in $L$, the momentum
dependence of the mode is irrelevant and all modes give the same contribution.

Close to the transition, $\xi\to \infty$, $ r_0\to\infty$. Taking the
limit $L\to\infty$ while keeping $ L/\xi$ finite leads to 
\begin{align}
\nu_{p} & \approx\cosh(2 r_0)\sqrt{\left(\frac{\sinh(L/\xi)}{L/\xi}\right)^{2}-1}.
\end{align}
We see again that the eigenvalue becomes independent of the momentum. 

Thus, we see that, in the large system size limit means that, for
the practical purpose of computing the EE away and close to the critical
point we may ignore the momentum dependence of the correlations. But
since the diagonal basis is related to the lab frame by an OBC Fourier
transform and two local squeezing transformation that do not affect
the entanglement this means that the contribution to the entanglement
in the GGE framework of a single mode is also the EE of a single spatial
site. Thus the two results are equivalent. 

Since we expect the GGE to hold for small system sizes $\frac{l}{N}\ll1$,
this extend our analytical results for the entanglement in this limit. 

We will now show that a similar statement holds in the reciprocal
phase. 

\paragraph{Reciprocal phase}

Recall the expressions for the correlations in the eigenbasis : 
\begin{align}
\overline{\langle\hat{b}_{p}^{\dagger}\hat{b}_{p}\rangle}= & \frac{1}{2}\left(\cosh\left(2 r_0\right)-1\right),\\
\overline{\langle\hat{b}_{\bar{p}}\hat{b}_{p}\rangle}= & \frac{1}{L}\sinh\left(2 r_0\right)\frac{2p^{2}\left(1-e^{i\varphi L}\right)}{\varphi\left(4p^{2}-\varphi^{2}\right)}.
\end{align}
The corresponding eigenvalue $\nu_{p}$ is given by 
\begin{equation}
\nu_{p}=\sqrt{1+\sinh^{2}2 r_0\left(1-\frac{32p^{4}\left(1-\cos\left(\varphi L\right)\right)}{\left(\varphi L\left(4p^{2}-\varphi^{2}\right)\right)^{2}}\right)}\label{eq:nurecp}
\end{equation}
Away from the transition $\varphi$ is finite and, in the large $L$
limit, $\overline{\langle\hat{b}_{\bar{p}}\hat{b}_{p}\rangle}\approx0$. 

Close to the transition $\varphi\to0$. Taking $L\to\infty$ and
keeping $\varphi L$ finite leads to : 
\begin{equation}
\overline{\langle\hat{b}_{\bar{p}}\hat{b}_{p}\rangle}\approx\frac{\sinh\left(2 r_0\right)}{2}\frac{\left(1-e^{i\varphi L}\right)}{\varphi L},
\end{equation}
which is again independent of $p$. We thus get that the GGE approach
is equivalent to the minimal bipartition in the reciprocal phase as
well. 

\subsection{Entanglement entropy and critical scaling}

\paragraph{Non-reciprocal phase}

In the non-reciprocal phase, the expression for $\nu_{n}$ is given
by (\ref{eq:nunonrecp}). 

Far from the critical point , we had $s(\nu_{p})\approx\ln\nu_{p}\approx rN,$
and thus 
\begin{equation}
S_{A}=rlN,
\end{equation}
which leads to the super-volume law scaling. 

To get the scaling near the critical point, we consider the limit
$L/\xi\ll1$ which corresponds to a regime where the localization
length is much greater than the system size. Expanding (\ref{eq:nunonrecp})
in powers of $L/\xi$ leads to 
\begin{equation}
\nu_{p}\approx N\frac{\Delta}{\sqrt{3}w}\left(1+\frac{1}{15}\left(L/\xi\right)^{2}\right)
\end{equation}
and 
\begin{equation}
S_{A}\approx l\left(\ln N+\frac{1}{15}\frac{\Delta^{2}-g^{2}}{w^{2}}N^{2}\right).
\end{equation}

\paragraph{Reciprocal phase}

The eigenvalue $\nu_{p}$ in the reciprocal phase is given by (\ref{eq:nurecp}).

Away from the transition, in the large $N$ limit, we have $\overline{\langle\hat{b}_{\bar{p}}\hat{b}_{p}\rangle}=0$
so the EE is simply 
\begin{equation}
S_{A}\approx ls\left(\cosh\left(2 r_0\right)\right).
\end{equation}
Close to the transition, in the continuous limit, and for $\varphi L\ll1$,
we obtain 
\begin{equation}
\nu_{n}\approx\frac{\Delta N}{\sqrt{3}w}\left(1+\frac{1}{15}\left(L/\xi\right)^{2}\right)
\end{equation}
which yields the same critical scaling than the non-reciprocal phase
:
\begin{equation}
S_{A}\approx l\left(\ln N+\frac{1}{15}\frac{\Delta^{2}-g^{2}}{w^{2}}N^{2}\right).
\end{equation}

\section{Local squeezing plus thermal occupation ansatz}\label{app:sqplth_ansatz}

In this appendix, we show that simply considering particle number is insufficient to fully characterize the spatial profile of entanglement. 

\begin{figure}[hptb]
    \centering
    \includegraphics[width=\columnwidth]{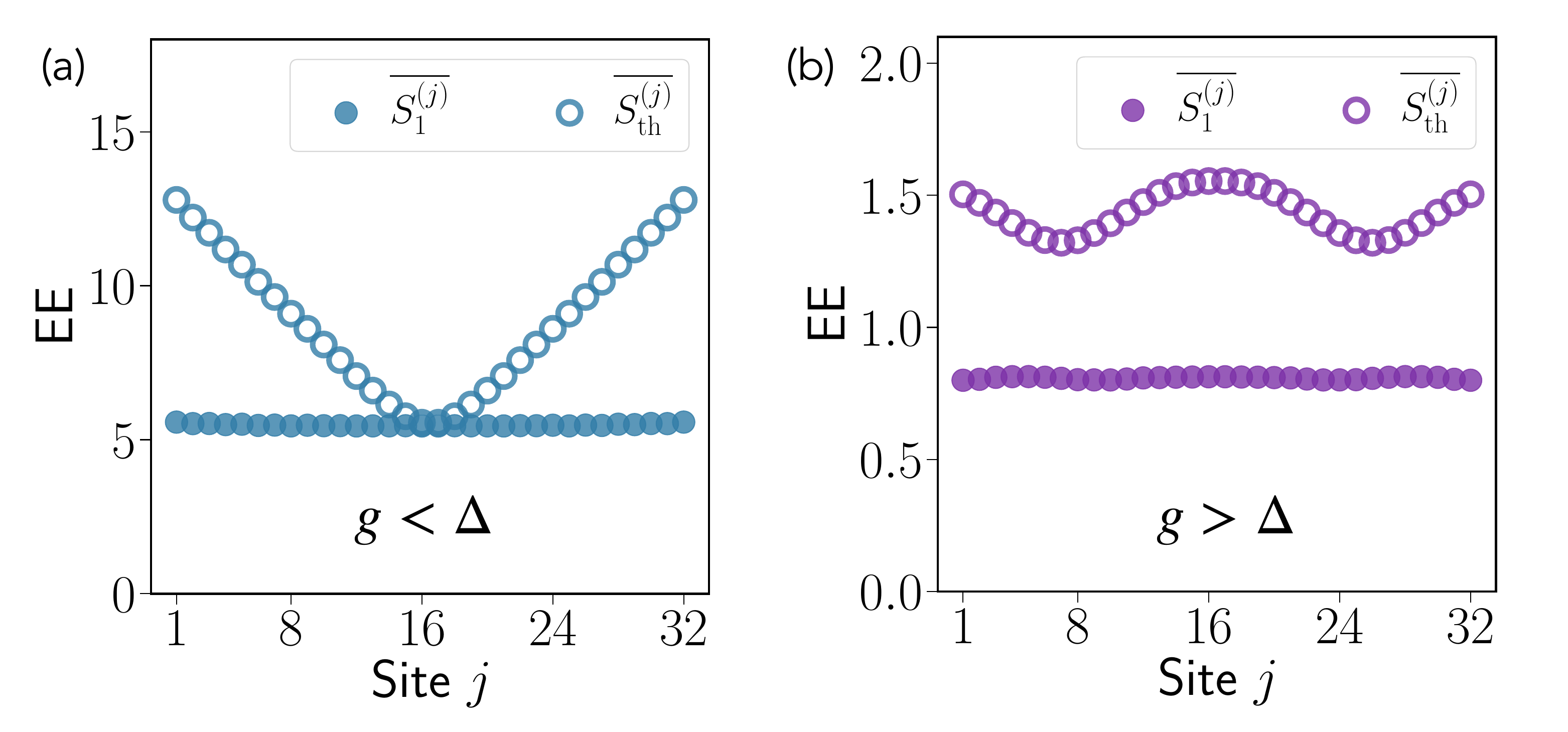}
    \caption{Comparison of expected thermal entropy $\overline{S_{\rm th}^{(j)}}$ against actual EE $S_1^{(j)}$ for a chain of size $N=32$ in \textbf{a.} the non-reciprocal phase, with $w = 1, g = 0, \Delta = 0.25$ and \textbf{b.} the reciprocal phase, with $w = 1, g = 0.3, \Delta = 0.25$. }
    \label{app_fig:thermal_mistake_plots}
\end{figure}

Given only access to the density profile, what kind of EE profile might one expect such a system to have? One reasonable approach would be to think about particle number as a proxy for the size of the local Hilbert space, and in general we expect a larger local Hilbert space to indicate that the site is more entangled with the rest of the system. To make this concrete, suppose a site $j$ has occupation $\langle \hat{a}_j^{\dag}\hat{a}_j \rangle$. One can try to associate an entropy with this density by assuming that when the system has thermalized, the density matrix of the site $j$ will be approximately that of a thermal state, for which the entanglement entropy is given by \begin{equation}
    S_{\rm th}^{(j)} = (\langle \hat{a}_j^{\dag}\hat{a}_j \rangle + 1) \ln (\langle \hat{a}_j^{\dag}\hat{a}_j \rangle + 1) - \langle \hat{a}_j^{\dag}\hat{a}_j \rangle \ln \langle \hat{a}_j^{\dag}\hat{a}_j \rangle,
\end{equation}
and use this as an estimate for the actual EE $S_1^{(j)}$. We compare these two quantities by plotting their time-averaged values against each other in both the reciprocal and non-reciprocal phases (Fig.~\ref{app_fig:thermal_mistake_plots}). In both cases, true entanglement profile is flat up to finite-sized effects, whereas the expected thermal entropy reflects localization in the non-reciprocal case and periodic spatial oscillations in the reciprocal case. In both cases, the thermal entropy significantly overestimates the true EE.

Another way to appreciate the relationship between particle number and entanglement is to explicitly extract the local squeezing and temperature of the time-averaged one-site density matrix. To do so, note that any diagonal single-site covariance matrix $\sigma$ can be decomposed into a rotations followed by a squeezing operation on a thermal state:
\begin{equation}
    \sigma = 
    R \begin{pmatrix}
        e^{z} & 0 \\ 0 & e^{-z}
    \end{pmatrix} \begin{pmatrix}
        e^{2 \beta} & 0 \\ 0 & e^{2 \beta}
    \end{pmatrix} \begin{pmatrix}
        e^{z} & 0 \\ 0 & e^{-z}
    \end{pmatrix} R^T,
\end{equation}
where $R$ is some orthogonal matrix and $e^{2 \beta \pm 2 z}$ are the eigenvalues of $\sigma$. The symplectic eigenvalue is entirely determined by $\beta$ and the local squeezing parameter $z$ does not affect entanglement properties at all. The quantities $\beta, z$ can be easily obtained by diagonalizing $\sigma$. 

\begin{figure}[hptb]
    \centering
    \includegraphics[width=\columnwidth]{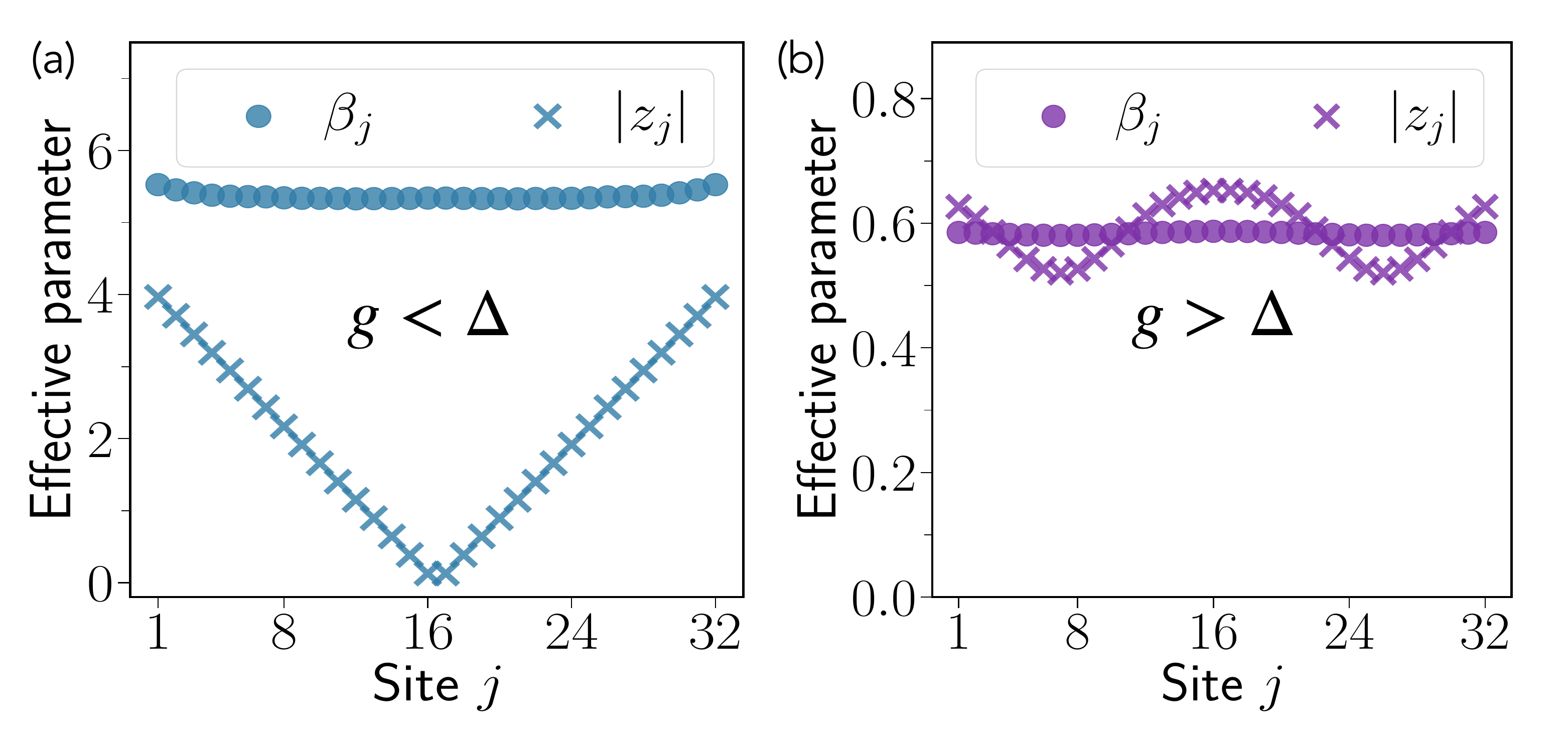}
    \caption{Comparison of local squeezing $|z_j|$ and local temperature $|\beta_j|$ for for a chain of size $N=32$ in \textbf{a.} the non-reciprocal phase, with $w = 1, g = 0, \Delta = 0.25$ and \textbf{b.} the reciprocal phase, with $w = 1, g = 0.3, \Delta = 0.25$. }
    \label{fig:beta_v_z}
\end{figure}

In Fig.~\ref{fig:beta_v_z}, we plot the values $\beta_j, z_j$ extracted from the time-averaged covariance matrix $\sigma_j$ at site $j$. These quantities display qualitatively similar features to the entropies of the previous plot, with temperature being analogous to EE and local squeezing displaying the same spatial distribution as the particle number. Notably, 
this allows us to make the more explicit statement that the spatial non-uniformity arising from non-reciprocity can be entirely characterized by local squeezing operations, which do not affect entanglement properties.

\section{Two enabling claims in the nonreciprocal phase}\label{app:4pt}

\subsection{The symplectic eigenvalue squared can be approximated using two point correlators}

In the main text, we assumed that in the nonreciprocal phase,
\begin{equation}\label{app_eq:eps4}
\begin{aligned}
    \varepsilon_4 := \frac{\overline{(\langle \hat{d}^{\dag} \hat{d} \rangle)^2} - \overline{\langle \hat{d}^{\dag} \hat{d}^{\dag} \rangle \langle \hat{d} \hat{d} \rangle} }{\left( \overline{\langle \hat{d}^{\dag} \hat{d} \rangle} \right)^2 - \left(\overline{\langle \hat{d}^{\dag} \hat{d}^{\dag} \rangle} \right) \left( \overline{\langle \hat{d} \hat{d} \rangle} \right) } - 1 \sim O \left(\frac{1}{N}\right).
\end{aligned}
\end{equation}
To show this rigorously requires a long and tedious calculation, which we will outline here. We will show this in two limits: (1) deep in the nonreciprocal phase, where we fix $r$ and take $N$ to be large, and (2) near the critical point, where we fix $N$ and take $g \rightarrow \Delta^-$. 

First, let us set up the problem and classify various conserved quantities. For simplicity, we will pick $\hat{d}$ to be the operator on the first site of the chain, in the frame where $j_0 = (N + 1)/2$ (picking any other site does not materially change the calculation, as we will see shortly). For brevity, we will denote 
\begin{equation*}
    t_s(x) := \sin\left(\frac{\pi x}{N + 1}\right)
\end{equation*} In this frame, the initial values of the momentum correlations are 
\begin{equation}
    \begin{aligned}
        \langle \hat{b}_k^{\dag} \hat{b}_q\rangle_0
        &= \frac{2}{N+1} \sum_{n=1}^N t_s(kn) t_s(qn) H(r, n,  r_0), \\
        \langle \hat{b}_k \hat{b}_q\rangle_0
        &= \frac{2}{N+1} \sum_{n=1}^N t_s(kn) t_s(qn) G(r, n,  r_0) (-1)^n,
    \end{aligned}
\end{equation}    
where we have defined the quantities 
    \begin{equation}
    \begin{aligned}
        &H(r, n,  r_0) \\
        &:= \frac{1}{2}\cosh \left( 2r \left( n - \frac{N+1}{2}\right) \right) \cosh(2 r_0) - \frac{1}{2}, \\
        &G(r, n,  r_0) \\
        & := -\frac{1}{2}\sinh \left( 2r \left( n - \frac{N+1}{2}\right) \right) \cosh(2 r_0) + \frac{i}{2} \sinh(2 r_0),
    \end{aligned}
\end{equation}
associated with the initial real space correlations in the tight-binding frame (see Eq.~(\ref{app_eq:initial_TB_real_space_corr})). This gives the full time-evolution of the four point functions as 
\begin{widetext}
    \begin{equation}
        \begin{aligned}
            \langle \hat{d}^{\dag} \hat{d} \rangle^2 &= \left( \frac{2}{N+1} \right)^2 \sum_{k, q, k', q' = 1}^N t_s(q) t_s(q') t_s(k) t_s(k') \langle \hat{b}_q^{\dag} \hat{b}_k \rangle_0 \langle \hat{b}_{q'}^{\dag} \hat{b}_{k'}\rangle_0 e^{i (\varepsilon_q + \varepsilon_{q'} - \varepsilon_k - \varepsilon_{k'})t}, \\
            |\langle \hat{d} \hat{d} \rangle|^2 &= \left( \frac{2}{N+1} \right)^2 \sum_{k, q, k', q' = 1}^N t_s(q) t_s(q') t_s(k) t_s(k') \langle \hat{b}_k \hat{b}_{k'} \rangle_0 \langle \hat{b}_{q}^{\dag} \hat{b}_{q'}^{\dag} \rangle_0 e^{i (\varepsilon_q + \varepsilon_{q'} - \varepsilon_k - \varepsilon_{k'})t},
        \end{aligned}
    \end{equation}
\end{widetext}
with $\varepsilon_k = \sqrt{w^2 + g^2 - \Delta^2} \cos \left( \frac{\pi k}{ N + 1}\right)$.

To calculate the time-average of the above quantities, we first observe that we can classify the conserved quantities into three sets $A, B, C$, defined by the conditions \begin{equation}
    \begin{aligned}
        &A: & q = k; && q' = k', \\
        &B: & q' = N+1-q; && k' = N+1-k, \\
        &C: & q = k'; && q' = k. \\
    \end{aligned}
\end{equation}
Note that for generic $N$, we always have this set of conserved quantities such that $\varepsilon_q + \varepsilon_{q'} - \varepsilon_k - \varepsilon_{k'} = 0$. For a specific values of $N$, we can have other sets of conserved quantities associated with special values of the cosine, but the effects of these do not scale with $N$, so we can ignore them. 

We define the sum over the conserved quantities in each set by $I^{r}_{A, B, C}, I^{a}_{A, B, C}$ for $\langle \hat{d}^{\dag} \hat{d} \rangle^2_t, |\langle \hat{d} \hat{d} \rangle_t|^2$ respectively, eg. 
    \begin{equation}
        \begin{aligned}
            I_A^r :=
            \left( \frac{2}{N+1} \right)^2 \sum_{k, q  = 1}^N t_s(q)^2 t_s(k)^2 \langle \hat{b}_q^{\dag} \hat{b}_q \rangle_0 \langle \hat{b}_k^{\dag} \hat{b}_k \rangle_0,
        \end{aligned}
    \end{equation}
and so forth. We note that these conserved quantities have some degeneracy between them, i.e. the sets $A, B, C$ have some overlap, for instance, $A$ and $B$ overlap when $q = k = N + 1 - k' = N+ 1 - q'$. In the continuous limit, these overlaps disappear. However, for any finite $N$, they are responsible for the $O(1/N)$ correction observable in the numerics. 

Finally, we observe that  \begin{equation}
    \begin{aligned}
        I^{r}_A &= \left( \overline{\langle \hat{d}^{\dag} \hat{d} \rangle_t}\right)^2, \\
        I^{a}_B &= \left( \overline{\langle \hat{d} \hat{d} \rangle} \right) \left( \overline{\langle \hat{d}^{\dag} \hat{d}^{\dag} \rangle} \right)= 0,
    \end{aligned}
\end{equation}
and 
\begin{equation}\begin{aligned}
    I_B^r &= I_C^r, \\
    I_A^a &= I_C^a.
\end{aligned}
\end{equation}
Hence, the time-average of the four-point functions reduces to 
\begin{equation}
    \begin{aligned}
        \overline{\langle \hat{d}^{\dag} \hat{d} \rangle^2_t} &= \left( \overline{\langle \hat{d}^{\dag} \hat{d} \rangle_t}\right)^2 + 2 I^{r}_B + O\left(\frac{1}{N} \left( \overline{\langle \hat{d}^{\dag} \hat{d} \rangle^2 }\right) \right), \\ 
        \overline{|\langle \hat{d} \hat{d} \rangle|^2} &= 2I^{a}_A +  O\left(\frac{1}{N} \left( \overline{| \langle \hat{d} \hat{d} \rangle |^2 }\right) \right)
    \end{aligned}
\end{equation}

With the problem set up, we can now do some computation. 

\subsubsection{Large $N$ limit}

First, note that in this case, $\left( \overline{\langle \hat{d}^{\dag} \hat{d} \rangle_t}\right)^2 \sim O (e^{2Nr} / N^2)$. We simply need to show that the correction due to $I_B^r - I_A^a$ is negligible compared to $\left( \overline{\langle \hat{d}^{\dag} \hat{d} \rangle_t}\right)^2$.


Generically, we might expect (and can show) that $I^{r}_B, I^{a}_A \sim O\left( e^{2Nr}/N^2 \right)$. This means that $ \left( \overline{\langle \hat{d}^{\dag} \hat{d} \rangle_t}\right)^2$ would be a pretty bad approximation for $\overline{\langle \hat{d}^{\dag} \hat{d} \rangle^2_t}$. Fortunately for us, it turns out that the differences will cancel in exactly the right way to allow us to make the desired approximation. First, let us write out the term in full: 
\begin{widetext}
     \begin{equation}
    \begin{aligned}
        &I^{r}_B - I^{a}_A \\
        &= \left( \frac{2}{N+1} \right)^4 \sum_{k, q = 1}^N t_s(q)^2 t_s(k)^2 \sum_{n, l = 1}^N t_s(qn) t_s(ql) t_s(kn) t_s(kl)  \left( H(r, l,  r_0) H(r, n,  r_0) - G(r, l,  r_0) G^*(r, n,  r_0) (-1)^{n + l}\right) \\
         &= \left( \frac{2}{N+1} \right)^2 \sum_{n, l} A(n, l) A(n, l) \left( H(r, l,  r_0) H(r, n,  r_0) - G(r, l,  r_0) G^*(r, n,  r_0) (-1)^{n + l}\right),
    \end{aligned}
\end{equation}
\end{widetext}
where we have defined \begin{equation}
\begin{aligned}
    A(n, l) = \frac{2}{N+1} \sum_{k = 1}^N t_s(k)^2 t_s(kl) t_s(kn),
\end{aligned}
\end{equation}
which can be solved explicitly to obtain \begin{equation}
    A(n, l) = \begin{cases}
        \frac{1}{2} + \frac{1}{4} \delta_{n, 1} + \frac{1}{4} \delta_{n, N}, & n = l \\
        - \frac{1}{4}, & | n - l| = 2  \\
        0, & \text{otherwise}.
    \end{cases}
\end{equation}
To show that $I_B^r - I_A^a$ is small, it suffices to pay attention only to the sum over $n = l$. Note that the $\delta_{n, 1}, \delta_{n, N}$ terms and the sum over $|n - l| = 2$ lead to a correction on the same order as the sum over $n = l$, as its contribution can be bounded by the former sum. Furthermore, the same calculation for sites other than the first site only change the condition to $|n - l| = 2k \quad {\rm mod} \quad (N + 1)$, where $k$ is the site number. Finally, we have 
\begin{widetext}
     \begin{equation}\label{app_eq:IB-IA}
    \begin{aligned}
        I^{r}_B - I^{a}_A &\sim \left( \frac{2}{N+1} \right)^2 \sum_{n = 1}^N
         \left( H(r, n,  r_0)^2  - |G(r, n,  r_0)|^2 \right) \\
         &= \frac{1}{2} \left( \frac{2}{N+1} \right)^2 \sum_{n = 1}^N \left( 1 - \cosh(2  r_0) \cosh \left(2r \left(n - \frac{N+1}{2}\right) \right) \right) \\
         &= \frac{1}{2} \left( \frac{2}{N+1} \right)^2 \left( N - \cosh(2  r_0) \frac{\sinh{Nr}}{\sinh{r}} \right) \\
         &\sim O \left( \frac{1}{N^2} e^{Nr} \right).
    \end{aligned}
\end{equation}
\end{widetext}

With that, we conclude that \begin{widetext}
    \begin{equation}
        \begin{aligned}
            \overline{\langle \hat{d}^{\dag} \hat{d} \rangle^2} - \overline{|\langle \hat{d} \hat{d} \rangle|^2}
            &= \left( \overline{\langle \hat{d}^{\dag} \hat{d} \rangle_t}\right)^2 - \left( \overline{\langle \hat{d} \hat{d} \rangle} \right) \left( \overline{\langle \hat{d}^{\dag} \hat{d}^{\dag} \rangle}\right) + O \left(\frac{1}{N} \left( \overline{\langle \hat{d}^{\dag} \hat{d} \rangle^2} - \overline{|\langle \hat{d} \hat{d} \rangle|^2} \right)\right) \\ 
            &= \left( 1 + O \left(\frac{1}{N} \right)\right) \left( \left( \overline{\langle \hat{d}^{\dag} \hat{d} \rangle_t}\right)^2 - \left( \overline{\langle \hat{d} \hat{d} \rangle} \right) \left( \overline{\langle \hat{d}^{\dag} \hat{d}^{\dag} \rangle}\right) \right).
        \end{aligned}
    \end{equation}
\end{widetext}

\subsubsection{Near critical point}



Near the critical point, we need to check what happens to $I_B^r - I_A^a$ when $g \rightarrow \Delta$. To investigate this limit, we first take $r \rightarrow 0$, then express $r, r_0$ in terms of $g, \Delta, w$.  Again, we only need to check the $n = l$ terms, with the other terms giving corrections of the same order. In this case, expanding in $r$ and ignoring the parts that cancel with some $R$ terms, Eq.~\ref{app_eq:IB-IA} becomes \begin{equation}
\begin{aligned}
    &I_B^r - I_A^a \sim \\
&\frac{1}{3} N \cosh(2r_0) r^2 + O(r^3) \\
&\simeq O\left( N ( \Delta^2 -g^2)/w^2\right),
\end{aligned}
\end{equation}
which results in only a $O(1/N)$ correction to the calculation using the squares of two point functions, which go as $N^2 (\Delta^2 - g^2)/w^2$.

We note that the calculation proceeds almost identically on the reciprocal side of the critical point. For the reciprocal side, the term $\langle \hat{d}^{\dag} \hat{d} \rangle^2$ does not oscillate at all and so does furnish any  corrections to the symplectic eigenvalue, whereas the corrections from squaring the $|\langle \hat{d} \hat{d} \rangle|^2$ terms again goes as $O(1/N)$.

\begin{figure}[hptb]
\centering \includegraphics[width=0.5\textwidth]{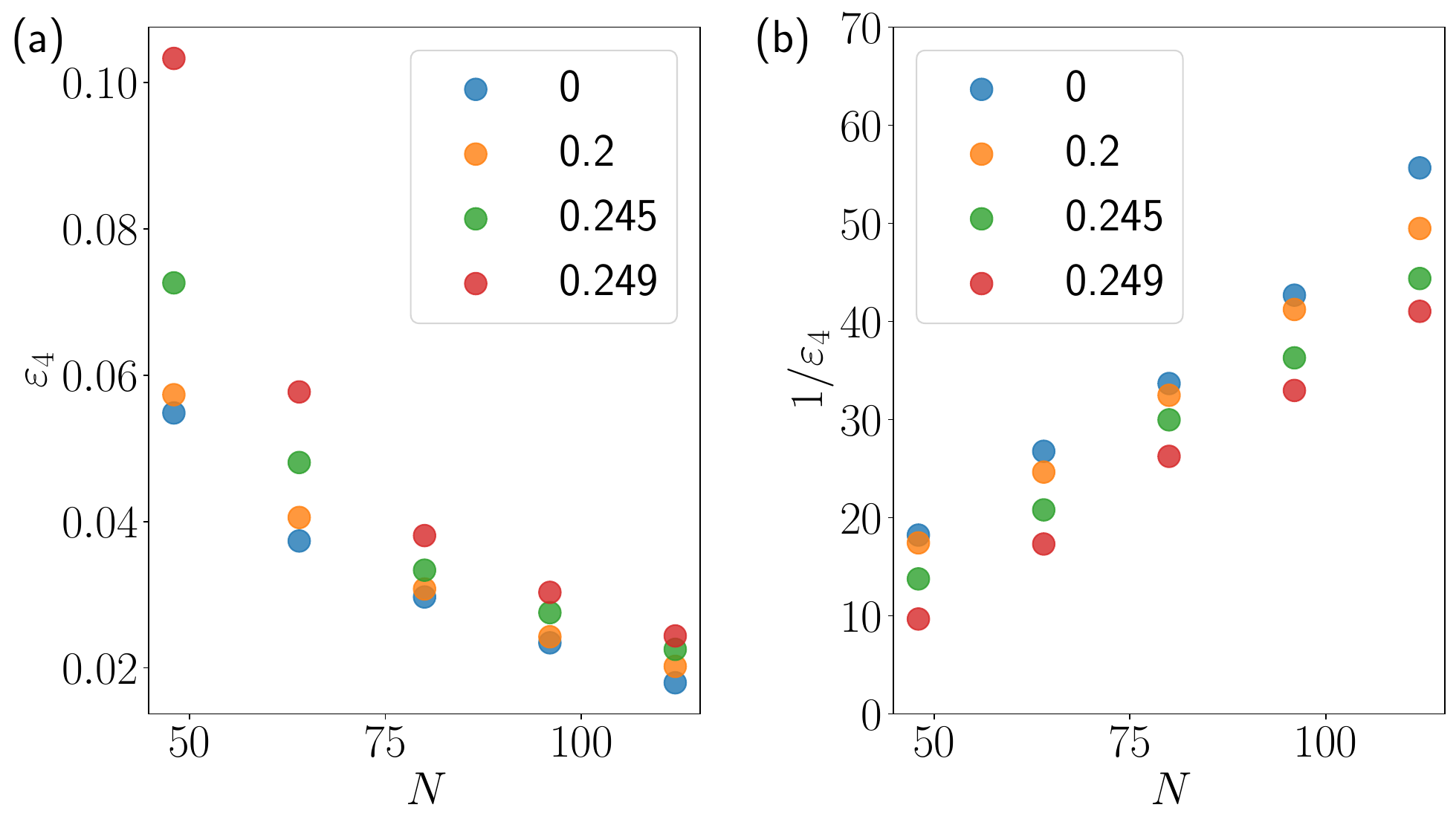} \caption{(a) Plot of the quantity $\varepsilon_4$ defined in Eq.~(\ref{app_eq:eps4}) for $w = 1, \Delta = 0.25, g = 0, 0.2, 0.245, 0.249$ and $N = 48, 64, 80, 96, 112$. (b) Plot of $1/\varepsilon_4$ for the same parameters. The behaviour is observed to be very clsoe to linear with $N$, demonstrating that indeed $\varepsilon_4 \sim O(1/N)$.}
\label{fig:eps4_plots} 
\end{figure}

To support these conclusions,  In Fig.~\ref{fig:eps4_plots}a, we plot the numerically calculated values of  $\varepsilon_4$ for the left-most site and observe that it does indeed go away as $N$ increases. We omit the $N=16,32$ points, as they do not fall neatly into the large $N$ regime, and we do not expect our assumptions to hold. In Fig.~\ref{fig:eps4_plots}b, we plot the inverse of the above quantity and observe that its behavior is indeed roughly linear with $N$. 

\subsection{Moving averages into the log only results in a small constant correction}

Next, in the main text, we also claimed that  
\begin{equation}\label{app_eq:S_t_first_correction}
    \frac{\overline{\nu_t^4 - (\overline{\nu_t^2}})^2}{(\overline{\nu_t^2})^2} \sim O(1).
\end{equation}
This allowed us to  take $\overline{\ln \nu_t^2} \simeq \ln \overline{\nu_t^2}$, since  $\ln \overline{\nu_t^2} \sim O(N)$. Now, $\nu_t^2$ involves taking products of four two point functions -- we will simply sketch the main ideas. 

First, note that the denominator goes as $O(e^{4Nr}/N^4)$. Now for each selection rule arising from taking a product of two point functions, one can take the most naive bound by simply taking absolute values of the summands and bounding the sine terms by $1$. In that case, one can show that a product of four two point functions goes at most as some the number of selection rules multiplied by $O(e^{4Nr}/N^4)$ as well. Since the number of selection rules is independent of $N$, we can consider it a constant. As such, all factors of $N$ in the numerator and denominator exactly cancel out. 

\begin{figure}[hptb]
\centering \includegraphics[width=0.5\textwidth]{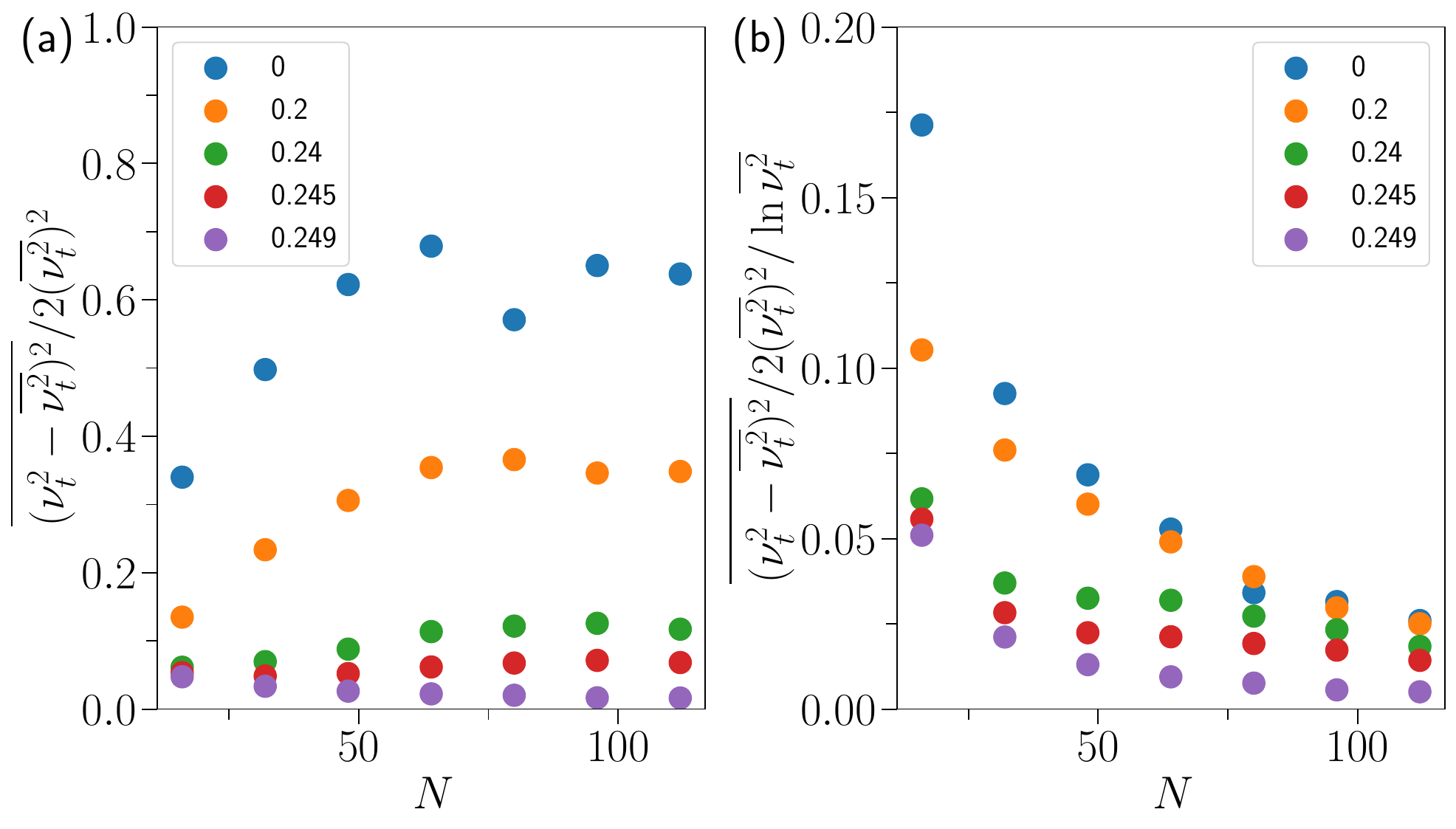} \caption{(a) Plot of the first correction to the estimate $\overline{\ln \nu_t^2} \simeq \ln \overline{\nu_t^2}$ in the non-reciprocal phase, with $w = 1, \Delta = 0.25, g = 0, 0.2, 0.24, 0.245, 0.249$. We observe that it appears to plateau as $N$ increases, consistent with the assumptions in the main text. (b) Plot of the first correction relative to the value of $\ln \overline{\nu_t^2}$, with the same parameters as in (a). We observe that the relative size of the correction appears to go to $0$ for large $N$.}
\label{fig:S_t_1_first_correction} 
\end{figure}

In Fig.~\ref{fig:S_t_1_first_correction}, we plot this quantity for the values of $N$ and $g \leq 0.25$ studied in the main text, and observe that indeed the correction to $\overline{\ln \nu_t^2}$ is exceedingly small compared to $\ln \overline{\nu_t^2}$, and approaches a constant for large $N$, as expected.

\end{appendix}
\end{document}